\documentclass[useAMS, usenatbib, usegraphicx]{mn2e}

\usepackage{amssymb}
\usepackage{amsmath} 
\usepackage{booktabs}
\newcommand\ApJ{ApJ}
\newcommand\MNRAS{MNRAS}
\newcommand\ARAA{ARA\&A}

\newcommand\PASJ{PASJ}

\newcommand\AnA{A\&A}

\def\alf{Alfv\'en\,}
\def\Gom{G\'omez\,}

\def\kon{K\"onigl }
\def\rud{R\"udiger}
\def\bq{\begin{equation}}
\def\eq{\end{equation}}

\let\grad=\nabla
\def\ds{\bmath{dS}}
\def\kz{\hat{k}_z}
\def\kr{\hat{k}_r}
\def\Bz{b_z}
\def\Bp{b_{\phi}}
\def\Jpa{\bmath{J_\parallel}}  % J_||
\def\vB{\bmath{v}_B}
\def\v{\bmath{v}} 
\def\gp{g^{\prime}}
\newcommand\cross{\bmath{\times}}

\newcommand\betaH{\tilde{\eta}_\mathrm{H}}
\newcommand\betaP{\tilde{\eta}_\mathrm{P}}
\newcommand\betaA{\tilde{\eta}_\mathrm{A}}
\newcommand\betaO{\tilde{\eta}_\mathrm{O}}
\newcommand\betaT{\tilde{\eta}_\mathrm{T}}

\newcommand\heta{\hat{\eta}}

\newcommand\hetaH{\hat{\eta}_\mathrm{H}}
\newcommand\hetaP{\hat{\eta}_\mathrm{P}}
\newcommand\hetaA{\hat{\eta}_\mathrm{A}}
\newcommand\hetaO{\hat{\eta}_\mathrm{O}}
\newcommand\hetaT{\hat{\eta}_\mathrm{T}}

\newcommand\bs{\tilde{\sigma}}
\newcommand\bk{\tilde{k}}
\newcommand\hk{\hat{k}}
\newcommand\hs{\hat{\sigma}}

\def\curl{{\grad \cross}}
\def\div #1 {\grad \cdot #1}

\def\v{\bmath{v}}
\def\k{\bmath{k}}
\def\hbk{\bmath{\hat{k}}}

\def\B{\bmath{B}}
\def\hB{\bmath{b}}
\def\E{\bmath{E}}            % E
\def\hs{\hat{\sigma}}

\def\oa{{\bmath{\omega}}_A}
\def\J{\bmath{J}}

\def\dv{\delta\v}
\def\hdv{\hat{\delta{\v}}}
\def\hdvr{\hat{\delta{v_r}}}
\def\hdvp{\hat{\delta{v_{\phi}}}}
\def\hdvB{\hat{\delta{\vB}}}
\def\hdvBr{\hat{\delta{v}_{_{Br}}}}
\def\hdvBp{\hat{\delta{v}_{_{B\phi}}}}

\def\dB{\bmath{\delta\B}}
\def\hdB{\bmath{\hat{\delta B}}}
\def\hdBr{\hat{\delta B}_r}
\def\hdBp{\hat{\delta B}_{\phi}}
\def\dBr{\delta B_r}
\def\dBp{\delta B_{\phi}}

\def\r{\bmath{r}}
\def\ph{\bmath{\phi}}

\newcommand{\delt} [1] {\frac{\partial #1}{\partial t}}

\def\Om{\bmath{\Omega}}

\def\hOm{\hat{\Om}}
\def\b1{{\bar{\omega}}}
\def\bo2{\bar{\omega}^2}
\title[Magnetorotational instability in diffusive discs]{Magnetorotational instability in magnetic diffusion dominated accretion discs}
\author[B.P.Pandey and Mark Wardle]
        {B.P. Pandey\thanks{E-mail:birendra.pandey@mq.edu.au; mark.wardle@mq.edua.au} and Mark Wardle \\
Department of Physics and Astronomy, Macquarie University, Sydney, NSW 2109, Australia}

\begin{document}

\date\today

\pagerange{\pageref{firstpage}--\pageref{lastpage}}

\pubyear{2012}

\maketitle

\label{firstpage}

\maketitle 
\begin{abstract}
We investigate the stability of partially ionised, differentially rotating, diffusive disc threaded by both azimuthal and vertical magnetic field. The general stability criterion of such a disc in the presence of axisymmetric fluctuations can be stated purely in terms of ambipolar and Hall diffusivities. It is shown that the disc is magnetorotationally unstable if the sum of scaled ambipolar and Hall diffusivities are larger than some numerical constant determined by the rotation profile of the disc. This criterion suggests that subthermal diffusive discs are always unstable to almost radial fluctuations. The field geometry and obliqueness of wavevector (encapsulated together in the topological factor $g$), plays dual role of not only assisting MRI in ambipolar dominated disc but also making otherwise stable region in Hall—-ambipolar diffusion plane unstable.

The vertical magnetic field and transverse \alf fluctuations are fundamental to magnetorotaional instability in Hall—-Ohm diffusion dominated discs. The disc having both azimuthal and vertical field in the presence of oblique axisymmetric fluctuations can be mapped to a purely vertical field threaded disc with purely vertical wavenumber. Although, azimuthal field and obliqueness of wavevector rescales the growth rate and window of instability, such a scaling permits MRI to operate much closer to the midplane in Hall—-Ohm dominated disc than is otherwise possible for a purely vertical field threaded disc.

The maximum MRI growth rate in the ambipolar diffusion dominated disc is proportional to the topological factor $g$. For long wavelength fluctuations the maximum growth rate is two fifth of ideal MHD case ($3/4\,\Omega$). The short wavelength fluctuations are likely to be damped by the diffusion. 

The diffusive discs are prone to new kind of MRI-–the diffusive MRI which has no counterpart in non—-diffusive discs. The diffusive MRI is caused by interplay between advection and diffusion of the field. However, since differential rotation of the disc is also at the heart of diffusive MRI, the maximum growth rate of the instability is identical to ideal MHD case. 

The excessively diffusive accretion discs in addition can also become unstable due to interplay between differential rotation and field diffusion. The ensuing diffusive instability grows at same maximum rate as MRI. However, the transport of angular momentum in the disc may not be efficient due to diffusion instability.
\end{abstract}


\section{Introduction}
It was Fred \cite{H60} who first suggested that only a magnetohydrodynamic model can explain the outward transport of angular momentum and resulting slow rotation of the protosolar nebula. Although, Hoyle$\textquoteright$s proposal was {\it well received} by the scientific community \citep{S72}, the detailed mechanism of  magnetic field--assisted angular momentum transport remained elusive for another three decades before \cite{BH91} discovered such a mechanism: the magnetorotational instability (MRI). Though the stability of magnetised Coutte flow  was analysed by \cite{V59} and \cite{C61}, their work remained largely unnoticed until \cite{BH91} not only clarified the underlying physical mechanism of this instability but by specifically applying it to accretion discs, showed that this instability will play an important role in angular momentum transport in the disc. Only after Balbus $\&$ Hawley$\textquoteright$s work did it become clear that a weak, subthermal magnetic field can facilitate the local transport of angular momentum in a weakly magnetised Keplerian disc. The most attractive feature of this instability is that the growth rate depends only on the differential rotation of the disc, an abundant source of free energy. The identification of this powerful local instability has provided a firm physical foundation for the parameterisation of viscosity in the turbulent discs \citep{SS73, HB91, S96, BH98, B03, B09}. 

The initial formulation of MRI by \cite{BH91} was for a fully ionised, {\it ideal} magnetohydrodynamic (MHD) medium in which field and fluid are well coupled to each other, a far cry from the solar nebula where often {\it non—ideal} MHD effects such as Ohm, Hall and ambipolar diffusion, play an important role. In fact, a large fraction of matter in the universe is not in a fully ionised state but is partially ionised with varying degree of ionisation determined mainly by local thermodynamic conditions. It is well known that in partially ionised discs viz. the solar nebula, protoplanetary discs, the discs around cataclysmic variables and black holes, the number of plasma particles, particularly away from thermal source, is often very small compared to the neutrals and thus, the plasma particles are not {\it frozen in} the magnetic field but can slip through it due to collisions with neutrals \citep{MS56}. This realisation led to the generalisation of MRI to weakly ionised discs \citep{BB94} where collision between the neutral and plasma particles are major source of energy dissipation. However, if the neutral-—ion collision time is faster than the rotation period of the disc, the medium moves as a single fluid and MRI in such a medium remains unaffected by collisions. In the opposite limit, when the neutral—-ion collision time becomes comparable or exceeds the rotation period, resistive quenching of MRI occurs in the disc \citep{M95, J96, HS98, SM99, T07, BS11}.
    
The magnetic field in a differentially rotating diffusive disc can evolve either (a) due to advection by the fluid (when diffusion is weak), or, (b) due to both advection and diffusion, or, (c) only due to diffusion (when advection by the fluid is negligible). Whereas advection dominated discs are described in the ideal MHD framework, remaining cases, (b) and (c) are important when {\it non-—ideal} MHD effects can not be neglected.  When (b) is valid, discs can become MRI unstable along one of the two possible pathways: (i) magnetic fluctuations are time dependent and its evolution is both due to diffusion and fluid advection \citep{W99, BT01, KB04, D04, WS11} and, (ii) magnetic fluctuations are almost time independent and field drift is exactly offset by the advection. We shall name this last case, diffusive MRI (DMRI) \citep{WS11}. However, when differentially rotating disc is {\it excessively} diffusive that is when diffusion dominates advection of the field, there is no MRI or, DMRI but a new kind of diffusive instability (DI) appears in the disc even when there is no fluid motion to transfer the angular momentum \citep{RK05, K08, BGB11, WS11}. Therefore, presence of favourable magnetic field topology in diffusive discs opens up new channels through which shear energy can be transferred to waves. 

The realisation that non--ideal MHD effects not only causes dissipation but can also assist MRI emerged only after it was shown that Hall effect, depending on the sign of magnetic helicity\footnote{sign of the projection of local magnetic field on the rotation axis}, can significantly modify the instability (Wardle, 1999, hereafter W99, Balbus \& Terquem, 2001, hereafter BT01). The role of ambipolar diffusion in assisting MRI was soon discovered \citep{KB04, D04}. It was shown that the ambipolar diffusion (AD) not only causes anisotropic dissipation, but under favourable field geometry can also assist the instability. More importantly, unlike ideal MHD where angular velocity must decrease outward in the disc, Hall and AD modified MRI is not dependent on the angular velocity profile \citep{BT01, KB04, D04}. Only the magnetic field topology and direction of wavevector becomes central to the MRI in the partially ionised discs \citep{KB04, D04} and role of angular velocity profile is relegated to insignificance. All in all, {\it non-ideal} MHD effects plays dual role in the disc: not only it causes dissipation, but can as well assist MRI in the presence of favourable magnetic field topology. 
  
Often, astrophysical discs are not uniformly ionised. The radial as well as vertical profile of the ionised matter depends on the proximity of the disc to the thermal and non-—thermal sources. For example, the surface layer of the protoplanetary discs (PPDs) is ionised by both cosmic \citep{UN81}, and x—-rays \citep{GNI97, IG99, F05, E08}. However, close to midplane, matter is largely shielded from external ionising sources and thus, except for the tiny ionisation by radionuclides \citep{UN09}, matter is mostly neutral. As a result vertical structure of the disc is layered, with surface layer well coupled to the magnetic field, whereas midplane of the disc, magnetically dead \citep{G96, S00, B03, SW03, SW08, TR08, BG09, TD09, TCS10, A11}. Clearly, the nature of magnetic diffusion through the disc changes with changing ionisation profile. In PPDs, Ohm--Hall is the main diffusion closer to the midplane whereas ambipolar—-Hall diffusion dominates the disc surface \citep{W07, B11}. Since MRI in such a disc can not operate closer to the midplane, it raises an interesting question: can local magnetic field topology at least extend the domain in which MRI operates, particularly closer to the disc midplane? As has been shown recently by Wardle \& Salmeron (2011; hereafter WS11) the answer to this question is in the affirmative. They showed that if the vertical field is aligned to the rotation axis, the MRI can be excited at much lower ionisation level than is otherwise possible. Therefore, ambient ionisation together with the local field topology determines the efficiency of MRI in transporting angular momentum particularly close to the midplane of PPDs.     

In the present work we investigate the MRI in a differentially rotating, partially ionised disc threaded by both azimuthal and vertical magnetic fields, i.e. $\B = (0, B_{\phi}, B_z)$. We shall assume that the fluctuations are axisymmetric and the wavevector $\k$ has radial and vertical components, i.e. $\k = (k_r, 0, k_z)$. It is pertinent to note here that the stability of such a diffusive disc was analysed by \cite{KB04} and \cite{D04}. Notwithstanding the simultaneous appearance of these works, the general dispersion relation was analysed only by \cite{D04} who not only provides a clear explanation of the underlying physical mechanism of MRI in ambipolar diffusion dominated discs but also identifies crucial topological parameter $g$ (which encapsulates disc field topology and obliqueness of wavevector) that assists the instability.  However, as we shall see, analytical results of \cite{D04} pertains to excessively diffusive discs, that is when fluid advection of the field is unimportant. Clearly, several important questions remain unanswered. For example, is it possible to state a general stability criterion for diffusive discs? What is the relative role of fluid advection and field diffusion in destabilising the disc? Does ambipolar diffusion affects the window over which Hall diffusion causes MRI? Can we identify the most MRI favoured field geometry in the disc? How does the maximum growth rate in various diffusive regimes compares with the ideal MHD case? How is the critical wavelength of the instability affected by the diffusion and the field? To clarify these issues, and to compare the behaviour of a magnetorotationally unstable disc in various diffusive regimes, we reinvestigate the stability of a diffusive disc in the present work. Thus, notwithstanding the similarity of our formulation with above mentioned works, not only the treatment of the problem but also the set of questions that we wish to address here are quite different.       

The paper is organised in the following fashion. In section 2 we briefly describe the basic set of equations suitable for the magnetised, non--self gravitating, diffusive disc rotating around a central point mass. In section 2.1 we give linearised equations in terms of diffusion tensor before describing general dispersion relation. By using a proper scaling, the dispersion relation is reduced to a very simple form which allows us to infer the general stability criterion in diffusion dominated discs. In following subsections, we show that diffusion plane, depending on most unstable and cut—-off wavenumbers can be subdivided in three distinct regions. In section 3, dispersion relation is analysed in two dimensional Hall-—Ohm (subsection 3.1) and Hall—-ambipolar (subsection 3.2)  diffusion planes. In section 4, limiting cases of dispersion relation is discussed. Section 5 contains several subsections discussing possible applications of results to PPDs. A brief summary of results is presented is section 6.   

\section{Formulation}
We shall start with standard description of a non self-—gravitating, differentially rotating, magnetised, thin disc of partially ionised matter which is threaded by both toroidal and poloidal magnetic field $\B = (0, B_{\phi}, B_z)$. We adopt cylindrical $\left(\mbox{r}\,, \phi\,, \mbox{z}\right)$ coordinate system anchored at the central star of mass $M$ and assume that the angular  velocity of the disc, $\Omega(r)$ is constant on the cylinder. The velocity of the fluid in the local frame corotating with the disc angular velocity $\Omega(r)$ can be written as $\v = \bmath{V} -– \v_{\phi}(r_0)\, \hat{\ph}$. Here $\v_{\phi}(r_0) = r\,\Omega(r_0)$, $\bmath{V}$ is the velocity in the laboratory frame and $\Omega(r_0)$ is fiducial angular velocity at  $r_0$. The time derivative in the local frame $\partial_t$ becomes $\partial_t + \Omega(r_0)\,\partial_{\phi}$  in the laboratory frame. The continuity and momentum equations are
\begin{equation}
\frac{\partial \rho}{\partial t} + \grad\cdot\left(\rho\,\v\right) = 0\,,
\label{eq:cont}
\end{equation}
\bq
\frac{\partial \v}{\partial t} + \left(\v\cdot \grad\right)\v -– 2\,\Omega\,v_{\phi}\,\hat{\r} + \frac{\chi^2}{2\,\Omega}\,v_r\,\hat{\ph} = - \frac{\nabla\,P}{\rho} + \frac{\J\cross\B}{c\,\rho} \,,
\label{eq:meq}
\eq
where the epicyclic frequency $\chi$ is
\bq
\chi^2 = 4\,\Omega^2 + \frac{d\,\Omega^2}{d\,\mbox{ln}R}\,.
\label{eq:epc}
\eq
Various terms in the momentum equation have their usual meaning. The induction equation is
\begin{eqnarray}
\delt \B = \curl\left[
\left(\v\cross\B\right) - \frac{4\,\pi\,\eta_O}{c}\,\J - \frac{4\,\pi\,\eta_H}{c}\,\J\cross\hB
\right. \nonumber\\
\left.
+ \frac{4\,\pi\eta_A}{c}\,
\left(\J\cross\hB\right)\cross\hB
\right]\,,
\label{eq:ind}
\end{eqnarray}
where  $\hB$ is the unit vector along the magnetic field. The expression for Hall, $\eta_H$, ambipolar, $\eta_A$ and, Ohm, $\eta_O$  for weakly--ionised ion-—electron--neutral plasma are \citep{K89, BT01} 
\bq
\eta_H = \frac{c\,B}{4\,\pi\,e\,n_e}\,,\nonumber\quad 
\eta_A = \frac{D\,B^2}{4\,\pi\,\rho\,\rho_i\gamma_i}\,,\nonumber
\eq
\bq
\eta_O = \left(\frac{c^2}{4\,\pi}\right)\frac{m_e\gamma_e\,\rho_n}{e^2\,n_e}\,,
\label{eq:ddf}
\eq
where
\bq
\gamma_j = \frac{<\sigma\,v>_j}{m_i + m_n}\,,
\eq 
and $<\sigma\,v>_j$ is the rate coefficient of collisional momentum exchange. Above diffusivities can be written in compact form as \citep{PW08}
\bq
\eta_H = \left(\frac{v_A^2}{\omega_H}\right)\,,
\eta_A = D\,\left(\frac{v_A^2}{\nu_{ni}}\right)\,,
\mbox{and}\,,
\eta_O = \beta_e^{-1}\,\eta_H\,,
\label{eq:ddf}
\eq
where $D = \rho_n / \rho$ is the ratio of the neutral mass density to the bulk density, $\rho = \rho_n + \rho_i$, $v_A = B / \sqrt{4\,\pi\,\rho}$ is the \alf velocity, $\beta_e = \omega_{ce} / \nu_{en}$ is the electron Hall parameter, which is a ratio of the electron cyclotron to electron–-neutral collision frequencies. The Hall frequency is $\omega_H = \rho_i\,\omega_{ci} / \rho$. Although, above diffusivity expressions are for a dustless medium, our analysis is quite general and is equally applicable to partially ionised medium containing dust. One could simply use a more general expressions for diffusivities, e.g. \cite{W07}. 
 
Current $\J$ is given by Amp\'ere's law,
\begin{equation}
    \J = \frac{c}{4\pi}\curl\B\,.
    \label{eq:Ampere}
\end{equation}

The role of magnetic field diffusion in assisting or inhibiting MRI can be made transparent if, following WS11, we recast the induction Eq.~(\ref{eq:ind})  as
\bq
\delt \B = \curl\left[
\left(\v + \vB\right)\cross\B - \frac {4\,\pi\,\eta_O}{c}\,\Jpa\right]\,,
\label{eq:ind1}
\eq
where the magnetic field diffusion or, drift velocity\footnote{We shall use diffusion and drift terms interchangeably  as they both mean same physical process.}
 \bq
\vB = \eta_P\,\frac{\left(\grad\cross\B\right)_{\perp}\cross\hB}{B} -– 
\eta_H\,\frac{\left(\grad\cross\B\right)_{\perp}}{B}\,, 
\label{eq:md0}
\eq
and $\eta_P = \eta_O + \eta_A$ is the Pedersen diffusivity. The parallel and perpendicular components of the current in the preceding equations refer to the orientation with respect to the background magnetic field. Note that first term on the right hand side of Eq.~(\ref{eq:ind1}), in the bracket $\curl\v\cross \B$ is the usual advection term. In the absence of diffusion, when $\v_B = 0$ and $\eta_O = 0$ field is advected by the fluid at finite velocity $\v$. The second term $\curl\vB\cross \B$ in the bracket is due to slippage or, drift of the field lines through the fluid at finite velocity $\vB$. In the fluid frame, the magnetic drift velocity $\vB$ contains the effect of field diffusion.

In the absence of parallel current, i.e. setting last term to zero in Eq.~(\ref{eq:ind1}), the induction equation reduces to familiar form except now unlike ideal MHD, flux $\Phi = \int{\B\cdot d\bmath{S}}$ through the arbitrary surface $S$ encircled by a contour $C$ in the plasma is {\it not frozen in the fluid frame but in a frame moving with  $\v + \vB$}. Such a formulation of induction equation allows us to visualise magnetic field as {\it real physical entity} that is drifting through the fluid with a given velocity. In this {\it frozen--in} frame, with the field and fluid moving  at $\v + \vB$, we have $c\,\E_{\perp} + \left(\v + \vB\right)\cross\B = 0$, which in the presence of 
$\left(\curl\B\right)_{\parallel}$ becomes
\bq
\frac{d\Phi}{dt} = \eta\,\int_S\,\left(\curl\B\right)_{\parallel}\,\cdot\,\ds\,.
\label{eq:fx1}
\eq
It is clear from Eq.~(\ref{eq:fx1}) that the presence of parallel current and not Ohm diffusion is responsible for the flux non—conservation in a weakly ionised plasma. Since energy dissipation is inevitable in such collisional plasmas, at first sight this result appears paradoxical. The non—-ideal MHD effects, particularly Pedersen (Ohm + Ambipolar) diffusion, owing to its dissipative nature, must affect magnetic flux. However, energy loss comes from the redistribution or, relaxation of magnetic field and not from its annihilation (e.g. magnetic reconnection) \citep{P63}. Therefore, total flux is conserved even in the presence of Ohm diffusion only if parallel current is absent in the medium.     

\subsection{Dispersion relation}
We shall assume a thin differentially rotating disc with $v\,, c_s\,, v_A \ll v_K $ which is threaded by a magnetic field having both azimuthal and vertical components, i.e. $B = (0, B_{\phi}, B_z)$. Further, a general rotation profile $\Omega (r) \propto r^{- q}$ is assumed. The radial variation in the physical quantities is neglected by assuming $k_r\,r \gg 1$. Although non—-axisymmetric fluctuations can play an important role in magnetic field amplification in the disc \citep{BH92a, TP96}, our analysis is limited to axisymmetric fluctuations only.  Thus, linearising Eqs.~(\ref{eq:cont}) -- (\ref{eq:ind}), and assuming axisymmetric perturbations of the form $\exp \left( i\,\k \cdot {\bf x} + \sigma\,t\right)$, with $k = (k_r\,, 0\,, k_z)$, the continuity equation (in the Boussinesq approximation) becomes 
\bq
\k \cdot \dv = 0\,.
\label{eq:cml}
\eq

The $\left(r\,,\phi\right)$ components of  the momentum equation becomes
\bq
\hdv
 = \frac{i\,\hk\,\mu}{\left[\hs^2 + 2\,\kz^2\,\left(2 - q \right)\right]}\,\left(
\begin{array}{cc} \hs   &  2\,\kz^2\\
                  \left(q - 2\right) & \hs
  \end{array}
\right)\,\hdB\,,
\label{eq:lin1}
\eq
and magnetic field drift velocity can be written as
\bq
\hdvB 
 = \frac{i\,\hk}{\mu}\,\left(
\begin{array}{cc} \Bz^2\,\hetaP   &  s\,\hetaH + g\,\hetaP\\
                  - \Bz^3\,\hetaH & \mu^2\,\hetaP - g\,\Bz\,\hetaH 
  \end{array}
\right)\,\hdB\,,
\label{eq:fd1}
\eq
where $\hdv = \dv / v_A$, $\hdvB  = \dv_B / v_A$, $\hdB = \dB / B$, $\hk = k\,v_A / \Omega$, $\hs = \sigma / \Omega$, $\hB = \B / B$, $\kz = k_z / k$, $\mu = \left(\hbk \cdot \hB\right) \equiv \kz\,\Bz$ and $\hat{\bmath{\eta}} = \bmath{\eta}\,\Omega / v_A^2$. The topological factors $g$ and helicity $s$ are
\bq
g = - \kr\,\kz\,\Bp\,\Bz\,,\quad
s = \mu\,\kz \equiv \left(\hk \cdot \hB\right)\, \left(\hk \cdot \hOm\right)\,.
\eq 
Here $\hOm = \Om / \Omega$. In table~1 we provide the list of important symbols and their usual meaning to ease reading. The important role of geometric factor $g$ in assisting AD modified MRI has been noted by \cite{D04}. We shall see that not only maximum growth rate but also parameter window of the AD assisted MRI depends on this geometric factor.

\begin{table*}
\begin{minipage}{110mm}
\caption{List of frequently used symbols.}
\label{mathmode}
\begin{tabular}{@{}llll}
Sybmol & Explanation & Symbol & Explanation\\
\hline
$\v$  & advection velocity       & $\vB$     & Field drift velocity \\[2pt]    
$\bmath{V}$             & Velocity        &   $\kr$   & $k_r /k$   \\[2pt]
$\kz$                   & $k_z / k$        &   $\Bz$    & $B_z / B$  \\[2pt]
$\Bp$                   & $B_{\phi} / B$   &   $\mu $   & $\kz\,\Bz$  \\[2pt]
$\,s = \mu\,\kz$ &Helicity&$g = - \kr\,\kz\,\Bp\,\Bz$&Topological factor\\[2pt]
$\gp$    &    $g / \kz$          &        $\hB$     &   $\B / B$       \\[2pt]
$\hs$    &    $\sigma / \Omega$  &        $\bs$     &   $\hs / \kz$    \\[2pt]
$\hk$    & $k\,v_A / \Omega$     &        $\bk$     &   $\hk\,\Bz$     \\[2pt]
$\eta_P = \eta_O + \eta_A$  &  Pedersen Diffusivity & $\eta_T = \eta_O + \mu^2\,\eta_A$  & Total diffusivity\\[2pt]
$\hetaP\,\left(\hetaT\right)$ & $\eta_P\,\left(\eta_T\right)\,\Omega / v_A^2$    & $\betaP \,\left(\betaT\right)$    & $\hetaP\,\left(\hetaT\right) / \left(\mu\,\Bz\right)$ \\[2pt]  
 $\left(\hetaO\,, \hetaA\right)$& $\left(\eta_O\,,\eta_A\right)\,\Omega / v_A^2$
&$\left(\betaO\,,\betaA\right)$ & $\left(\hetaO\,,\hetaA\right) / \left(\mu\,\Bz\right)$\\[2pt]
       $\hetaH$    &
  $\eta_H \,\Omega / v_A^2$    & $\betaH$ & $\hetaH / \Bz$\\
\hline
\end{tabular}
\end{minipage}
\end{table*}
The linearised induction equation ($r\,,\phi$ components) becomes
\begin{eqnarray}
\left[
\left(
\begin{array}{cc} \hs   &  0\\
                   q & \hs
  \end{array}
\right) 
  + \frac{\hk^2\,\mu^2}{\left[\hs^2 + 2\,\kz^2\,\left(2 - q \right)\right]}\,\left(
\begin{array}{cc} \hs   &  2\,\kz^2\\
                  \left(q - 2\right) & \hs
  \end{array}
\right) \right.\,,\nonumber\\
\left. + k^2\,
\left(
\begin{array}{cc} \eta_{rr}   & \eta_{r\phi}\\
                   \eta_{\phi r} & \eta_{\phi\phi}
  \end{array}
\right)\,\right]\hdB = 0
\,,
\label{eq:lin2}
\end{eqnarray}
where the diffusivity tensor have following components 
\begin{eqnarray}
\eta_{r\,r} = \hetaO + \Bz^2\,\hetaA\,,\quad \eta_{r\,\phi} = s\,\hetaH + g\,\hetaA\,,\nonumber\\
\eta_{\phi\,r} = \left(g\,\hetaA –- s\,\hetaH\right) / \kz^2\,,  
\quad
\eta_{\phi\,\phi} = \hetaO + \left(1 -\kr^2\,\Bz^2\right)\,\hetaA\,.
\label{eq:amt}
\end{eqnarray}

The dispersion relation obtained from Eqs.~(\ref{eq:cml})-- (\ref{eq:lin2}) is
\bq
\hs^4 + \hk^2\,\left(\hetaP + \hetaT\right)\,\hs^3 + C_2\,\hs^2 + C_1\,\hs + C_0 =0\,.
\label{eq:mdr}
\eq 
where 
\begin{eqnarray}
C_2 = \hk^4\,\left(\hetaP\,\hetaT + \mu^2\,\hetaH^2\right)  - q\, \hk^2\,\left(s\,\hetaH + g\,\hetaA\right) 
\nonumber\\
+ 2\,\kz^2\,\left(2 - q\right) + 2\,\mu^2\,\hk^2\,,
\nonumber\\ 
C_1 = \hk^2\,\left[2\,\kz^2\,\left(2 - q\right) + \mu^2\,\hk^2\right]\,
\left(\hetaP + \hetaT\right)\,,
\nonumber\\
C_0 = 
\left[\mu^2\,\hk^2 + \left(\left(4 - q\right)\,s\hetaH –- q\,g\hetaA\right)\,\hk^2 –- 2\,q\,\kz^2\right] \,\mu^2\, \hk^2
\nonumber\\ 
 -– 2\,\kz^2\,\hk^2\,q\,\left(2 - q\right)\,\left(s\,\hetaH + g\,\hetaA\right)
\nonumber\\
+ 2\,\hk^4\,\kz^2\,\left(2 - q\right)\left(\hetaP\,\hetaT + \mu^2\,\hetaH^2\right)\,,
\label{eq:coefm}
\end{eqnarray}
and
\bq
\hetaT = \hetaP + \left(\mu^2 - 1 \right)\,\hetaA\,.
\eq
The dispersion relation, Eq.~(\ref{eq:mdr}) is well known in the literature \citep{KB04, D04}. However, as has been noted above, only \cite{D04} analysed the general dispersion relation.  We shall re--examine above dispersion relation de novo to answer following questions. What is the general criterion under which a diffusive disc is magnetorotationally unstable? Does ambipolar diffusion affect the domain over which Hall destabilises the disc? What is the role of $g$ in Hall diffusion driven MRI? How important is the role of vertical field and vertical wavevector in exciting MRI in diffusive discs? How does the presence of azimuthal field affect the instability?  Is unstable wavevector window a function of disc magnetic field? Does favourable field topology in a diffusive disc also implies that MRI can operate much closer to the midplane than the limit set by purely vertical field threaded disc (e.g. WS11)?  To reiterate, our analysis has a different focus and is complementary to \cite{D04}. 

Rescaling physical quantities as 
\begin{eqnarray}
\hs \rightarrow \bs = \hs / \kz\,,\quad 
\hk \rightarrow \bk = \hk\,\Bz\,, 
\hetaP \rightarrow \betaP = \frac{\hetaP }{\kz\,\Bz^2}\,,
\nonumber\\
\hetaT \rightarrow \betaT = \frac{\hetaT}{\kz\,\Bz^2}\,,\quad
\hetaH \rightarrow \betaH = \frac{\hetaH }{\Bz}\,,
\label{eq:scl}
\end{eqnarray}
the dispersion relation, Eq.~(\ref{eq:mdr}) can be recast in the following simple form
\bq
a\,\bk^4 + b\,\bk^2 + c = 0\,,
\label{eq:dnc}
\eq
where the coefficients $a$, $b$ and $c$ are
\bq
a = \beta - \gamma\,,\quad b = \delta\,\bs - \gamma\,\frac{c}{\bs^2}\,, 
\quad c = \bs^2\,\left[2\,\left(2 - q\right) + \bs^2\right]\,,
\label{eq:coeff1}
\eq
with 
\begin{eqnarray}
\beta = \left(\betaH^2 + \betaP\,\betaT\right)\,\bs^2 + \left(\betaP + \betaT\right)\,\bs 
\nonumber\\
+ 2\,\left(2 - q\right)\,\left[ \betaP\,\betaT 
+ \left(\betaH + \frac{1}{2 - q}\right)^2 \right]\,,\nonumber\\
\delta = \left(\betaP + \betaT\right)\,\left[2\,\left(2 - q\right) + \bs^2\right] + \left(\frac{4 - q}{2 - q}\right)\,\bs^2\,, 
\nonumber\\
\gamma = q\,\left(g^{\prime}\,\betaA + \betaH + \frac{1}{2 - q }\right)\,,
\label{eq:coeff2} 
\end{eqnarray}
and $g^{\prime} = g / \kz$. The dispersion relation, Eq.~(\ref{eq:dnc}) can also be recast as
\begin{eqnarray}
\beta\,\bk^4 + \delta\,\bk^2\,\bs + \left[2\,\left(2 - q\right) + \bs^2\right]\,\bs^2 = 
\nonumber\\
\gamma\,\bk^2\,\left[2\,\left(2 - q\right) + \bs^2 + \bk^2\right]\,,
\label{eq:dnc1}
\end{eqnarray}
which shows that for $q < 2$, coefficients $\delta$, $\beta$ and $\left[2\,\left(2 - q\right) + \bs^2\right]$ on the left hand side are all positive. Thus positive $\bs$ will require positive $\gamma$ as square bracket on the right hand side is always positive for $q < 2$. Therefore, a differentially rotating disc is unstable only if 
\bq
q\, \left(g^{\prime}\,\betaA + \betaH + \frac{1}{2 - q}\right) > 0\,.
\label{eq:pwd}
\eq
We have arrived at the most general stability criterion for a diffusive disc: whenever (i) $0 < q < 2$ and (ii) $\gp\,\betaA + \betaH + 1 / \left(2 - q\right) > 0$, a differentially rotating disc is unstable.  Note that Ohm diffusivity does not appear in the above expression which is not surprising considering that Eq.~(\ref{eq:pwd}) is the long wavelength limit of the sufficient stability condition, $C_0 < 0$.

The physical condition for the onset of instability can be presented as a competition between the fluid advection and magnetic field drift in the plasma. This can be seen from the azimuthal and radial components of Eqs.~(\ref{eq:lin1}) and (\ref{eq:fd1}) respectively, which suggest that when $\sigma = 0$, the radial advection of the fluid, $\hdvr$ is equal and opposite to  the radial drift velocity $\hdvBr$. This can be seen by combining azimuthal component of Eqs.~(\ref{eq:lin1}) and radial component of Eq.~(\ref{eq:fd1})] as 
\bq
\hdvr + \hdvBr = i\,k\,\mu\,\left[\frac{\gamma}{q}\,\hdBp + \frac{\Bz^2}{\kz}\,\betaA\,\hdBr \right]\,.
\label{eq:rbal}
\eq
Here we have neglected Ohm diffusion. By setting $\gamma = 0$ near marginal state, in the vicinity of $\eta_A \ll 1$, we get $\hdvr + \hdvBr \approx 0$. This provides a simple physical explanation for why diffusion will assist MRI. The outward radial slippage of the field due to diffusion weakens magnetic tension force. As a result magnetic restoring force in the wave gets dominated by the inertia force which causes increased inward radial drift of the fluid element. This is how diffusion assists MRI. 

As has been noted in the introduction, the magnetic diffusion opens up completely new channels through which shear energy can be transferred to waves. Therefore, it is natural to ask whether above stability criterion, Eq.~(\ref{eq:pwd}) could have been as well derived from reduced induction equation, i.e. dropping one or other term in Eq.~(\ref{eq:lin2}). It will be shown later (please see section 3) that in the limiting case, when advection and diffusion terms in Eq.~(\ref{eq:lin2}) balance each other,  we arrive at the identical criterion.       

For a diffusive Keplerian disc, since $q = 3/2$, condition (i) $q \in (0, 2)$ is automatically satisfied. Thus the stability criterion for a Keplerian disc becomes  
\bq
g^{\prime}\,\betaA + \betaH + 2 > 0\,,
 \label{eq:pwd1}
\eq
which states that the sum of ambipolar and Hall diffusivities with suitable coefficients determined by the obliqueness of wavevector and field topology should exceed some given numerical value predetermined by the Keplerian profile.  Eq.~(\ref{eq:pwd1}) provides a particularly simple stability test for diffusive discs. For example, Hall--Ohm dominated disc is unstable if $\eta_H > -2$. This stability criterion can be stated in terms of Hall and rotation frequencies as 
\bq
\Omega > - 2\,\Bz\,\omega_H \equiv – 2\,\omega_{HZ}.
\label{eq:CH1}
\eq
where $\omega_{HZ}$ is the Hall frequency defined in term of vertical field.   
For negative $\Bz$ the disc is magnetorotationally unstable if Hall frequency is less than one half of the Keplerian frequency. The Hall unstable region [e.g. Fig.~5, WS11 or, Fig.~1—-3, BT01] in a purely vertical field threaded disc ($\Bz = 1$) becomes arbitrarily small in $\Bz \rightarrow 0$ limit when the disc is threaded by non--vertical field. We see from Eq.~(\ref{eq:pwd1}) that at $\eta_H\,\Omega / v_A^2 = - 2\,\Bz$, since $\eta_A$ is always positive, ambipolar diffusion can assist MRI only if, $g > 0$, in agreement with \cite{D04}. 

From stability criterion Eq.~(\ref{eq:pwd1}) we see that whereas Hall--Ohm stable discs ($\eta_H \leq -2\,\Bz\,v_A^2 / \Omega$) may or may not be ambipolar unstable, Hall--Ohm unstable discs ($\eta_H > -2\,\Bz\,v_A^2 / \Omega$) can always become ambipolar unstable for non--vertical field and oblique wavevector since $g > 0$ can be easily satisfied. The ambipolar diffusion, on the other hand, not only assists MRI when $g > 0$, but also enlarges parameter space over which Hall can assist MRI. Therefore Hall assists MRI when 
\bq
\eta_H > - 2\,\Bz\,v_A^2 / \Omega –- \frac{g}{s}\,\eta_A\,.
\label{pwd2}
\eq   
Since $\Bz\,,\Bp \in [0,1]$ and maximum $g = 0.25$, above inequality implies that for non-zero $\eta_A$, Hall diffusion causes MRI at much larger negative value than for  purely vertical field (WS11, BT01). 

To summarise, stability criterion of a magnetised diffusive disc can be stated purely in terms of disc diffusivities. The MRI in the Hall--Ohm dominated disc is assisted by Hall diffusion if its value exceeds some numerical value predetermined by the disc rotation profile. In ambipolar--Ohm dominated discs, in addition, the topological factor $g$ (which carries the signature of non--vertical field and oblique wavevector) should be positive. The presence of such a positive topological factor in turn enlarges the parameter space over which Hall assists MRI. The topological factor $g$ plays the dual role: it not only assists MRI in ambipolar regime but also enlarges the domain of Hall assisted MRI.

\subsection{Division of diffusion plane}
It is clear from Eq.~(\ref{eq:dnc}) that the wavenumber range over which MRI occurs extends between zero and the cut—-off wavenumber 
\bq
\bk_c = \left[\frac{2\,\left(2 - q\right)\,\gamma}{2\,\left(2 - q\right)\,\left[\betaP\,\betaT 
+ \left(\betaH + \frac{1}{2 - q}\right)^2 \right] - \gamma}\right]^{1/2}\,.
\label{eq:kcc}
\eq  
The disc is magnetorotationally stable if $\gamma \leq 0$. In the opposite limit, when the disc is unstable, the diffusion space, like WS11 can be divided into three different regions depending on the denominator in Eq.~(\ref{eq:kcc}).  
\begin{figure}
\includegraphics[scale=0.32]{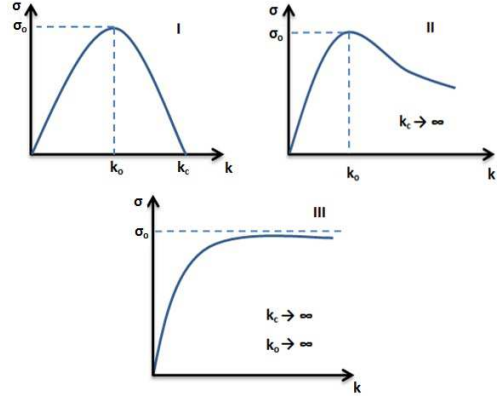}
\caption{The division of diffusion plane according to the values of wavenumbers $k_c\,,k_0$. Region I in diffusion plane correspond to finite  $k_c\,,k_0$. Region II correspond to finite  $k_c$ finite and $k_c \rightarrow \infty$ and region III correspond to  $\left(k_c\,,k_0\right) \rightarrow \infty$.}
\label{fig:FC}  
\end{figure}
When (a) $0 < \gamma < 2\,\left(2-q\right)\left[\betaP\,\betaT + \left(\betaH + \frac{1}{2-q}\right)^2\right]$, the cut—-off wavelength is finite [Fig.~(\ref{fig:FC}) labelled I]. On the other hand, when (b) $\gamma \geq 2\,\left(2-q\right)\left[\betaP\,\betaT + \left(\betaH + \frac{1}{2-q}\right)^2\right]$ the cut—-off wavenumber occurs at infinity. The unstable region in the diffusion plane in this case can be further subdivided depending on the value of most unstable wavenumber $k_0^2 = - b / 2\,a$ which is  
\bq
\bk_0^2 = \frac{\left[2\,\left(2-q\right) + \bs_0^2\right]\,\gamma - \delta\,\bs_0}{2\,\beta - \gamma}\,. 
\eq   
Here $\bs_0$ is maximum growth rate which can be derived by setting $b^2 -– 4\,a\,c = 0$ in Eq.~(\ref{eq:dnc}).

Thus unstable region when $k_c\rightarrow \infty$ is divided in region II 
and III depending on whether $\gamma > \delta\,\bs_0 / \left[2\,\left(2-\alpha\right) + \bs_0^2\right]$ or, $\gamma < \delta\,\bs_0 / \left[2\,\left(2-\alpha\right) + \bs_0^2\right] $ [Fig.~(\ref{fig:FC}) labelled II and III]. Region II correspond to finite $k_0$ and infinite $k_c$ whereas region III correspond to infinite $k_0$ and $k_c$. We shall utilise this analysis in the following section to examine the dispersion relation.

\section{Analysis of dispersion relation in two dimensional diffusion plane}
Coupling between the magnetic field and matter varies with height in protoplanetary discs. The detailed calculation of diffusivities at $1\,\mbox{AU}$ for a minimum mass solar nebula in which dust has settled to the midplane of the disc suggests that if the disc is threaded by a $\mu$G field, then Ohm is the dominant diffusion from midplane to three scale heights\footnote{which is determined by the hydrostatic balance between gravity and pressure gradient} at $1\,\mbox{AU}$  whereas Hall dominates between three and five scale heights \citep{W07, B11}. The ambipolar diffusion is dominant beyond five scale heights. The presence of dust does not change this picture either though at $5\,\mbox{AU}$, Hall becomes dominant diffusion very close to the midplane [Figs.~(10) and (11) in Wardle (2007) or Fig.~2 in Bai (2011)]. However, for stronger magnetic field [$B \in (0.5, 2)\,\mbox{Gauss}$], when dust has settled to the midplane, at $1 \mbox{AU}$ Hall is the dominant diffusion mechanism between midplane and $5$ scale heights. In the presence of $0.1\,\mu\mbox{m}$ dust, not Hall but Ohm becomes dominant between midplane and two scale heights \citep{W07, B11}. We may conclude that with changing scale height, as one moves from midplane to surface layer in PPDs, diffusivity typically changes from Ohm to ambipolar with Hall sandwiched in between, i.e. ($\eta_O \rightarrow \eta_H \rightarrow \eta_A$) with a large overlap region where Hall—-Ohm or,  Hall—-ambipolar is simultaneously important. Therefore, owing to the layered structure of PPDs it is important to carryout the analysis in two dimensional, Hall—-Ohm ($\eta_H-\eta_O$) or, Hall—-ambipolar ($\eta_H-\eta_A$) planes. Furthermore, such an analysis provides a clear physical interpretation of the dispersion relation.  To keep the analysis transparent we restrict ourselves to a Keplerian disc and assume $q = 3 / 2$. 

\subsection{Hall–-Ohm diffusion plane}
It is clear from above discussion that owing to its layered structure, Ohm and Hall is dominant diffusion close to the midplane of the protoplanetary disc. Therefore, we shall first analyse the dispersion relation, Eq.~(\ref{eq:dnc}) by retaining Ohm and Hall diffusion and set $\eta_A = 0$. The reduced dispersion relation in two dimensional diffusion planes provides a simple pictorial representation. This can be seen by writing the dispersion relation, Eq.~(\ref{eq:dnc}) as 
\bq
\left[\betaH + \frac{2}{1 + \bs^2} -– \frac{3\,A}{4}\right]^2 + \left(\betaO + \bs\,A\right)^2 = \left(\frac{3\,A}{4}\right)^2\,,
\label{eq:dno1}
\eq
where  
\bq 
A = \frac{1}{1 + \bs^2} + \frac{1}{\bk^2}\,.
\eq 
Eq.~(\ref{eq:dno1}) is identical to Eq.~(B10) of WS11 except now the physical quantities are normalised according to Eq.~(\ref{eq:scl}). Recall that in WS11, disc is threaded only by vertical magnetic field and wavevector is parallel to the magnetic field. That the dispersion relation derived for a disc with non-—vertical field and waves propagating at an oblique angle can be mapped to a purely vertical field with vertical wavenumber suggests that the vertical field and transverse fluctuations are fundamental to MRI in the Hall—-Ohm diffusion dominated discs. In the absence of either vertical field or, vertical wavenumber disc is stable to axisymmetric fluctuations. Therefore, we may conclude that the Hall—-Ohm dominated disc is magnetorotationally stable if waves are purely radial or, ambient field is purely azimuthal. We stress that this conclusion is valid only for the axisymmetric fluctuations. However, disc threaded by a purely toroidal field do become magnetorotationally unstable in the presence of non—axisymmetric fluctuations \citep{BH92a, TP96}.    

\begin{figure}
\includegraphics[scale=0.32]{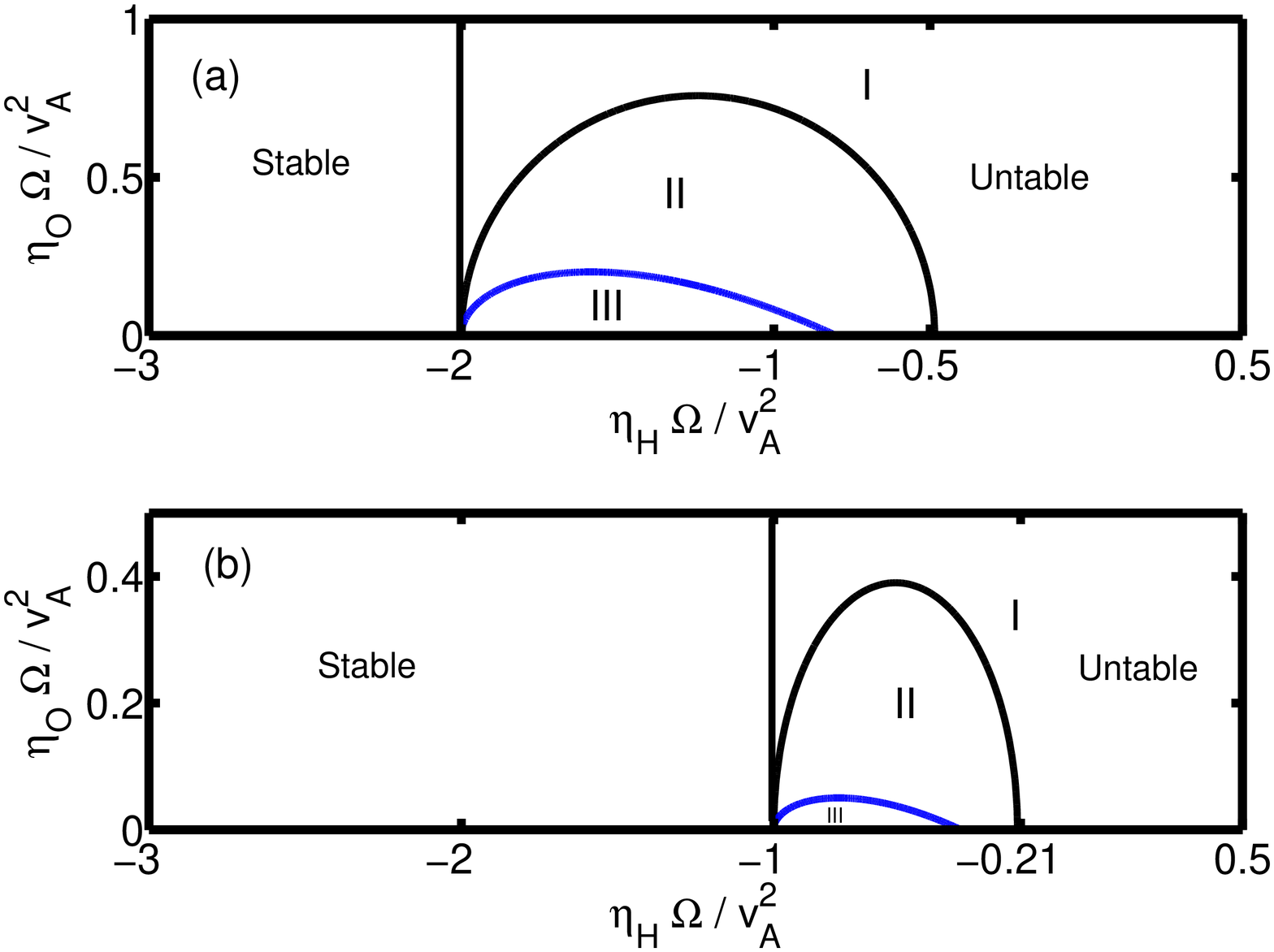}
\caption{The solution of Eq.~(\ref{eq:dno1}) in the Hall--Ohm diffusive plane is shown in the above figure. In Fig.~2(a), purely vertical magnetic field ($B_z = B$) and $\kz = 1$ is assumed whereas in Fig.~2(b), presence of both azimuthal and vertical fields ($B_z = 0.5\,B$) is assumed keeping $\kz$ unchanged. The vertical line divides stable region from magnetorotationally unstable region. The Hall assisted MRI region in the presence of purely vertical field [$\Bz = 1\,,  \hetaH > -2$, Fig.~2(a)] shifts to the right [$\hetaH > -2\,b_z = -1$, Fig.~2(b)] for non--vertical field.}
\label{fig:FO}  
\end{figure}

The above dispersion relation, Eq.~(\ref{eq:dno1}) describes a circular locus in $\eta_H - \eta_O$ plane for given $\sigma$ and $k$. Fig.~(\ref{fig:FO}) shows the dependence of the marginal locus on $\Bz$. Recall that magnetorotationally unstable region in a vertical field threaded disc with $\k = (0, 0, k_z)$ correspond to $\hetaH > -2$ [Eq.~(\ref{eq:pwd1}); see also Fig.~5 of WS11 or, Fig.~1--3 of BT01 where $\hk^2$ vs. $2\,\hetaH$ has been plotted]. Since in this case we see from Eq.~(\ref{eq:rbal})  that  $\hetaH > -2$ ($\Bz = \kz =1$) implies that Hall assists MRI only when the radial drift of the field exceeds radial advection of the field, i.e. $\hdvBr > \hdvr$. Thus, the instability interval $(-2, -1 / 2)$ in which all wavenumbers are unstable correspond to $\hdvBr > \hdvr$ and $\hdvBp  < \hdvp$ since  form Eqs.~(\ref{eq:lin1}) and (\ref{eq:fd1}) we get
\bq
\hdvp = - \frac{1}{2}\,i\,\hk\,\hdBr\,\quad\,\mbox{and}\,,\quad\hdvBp = - i\,\hk\,\hetaH\,\hdBr\,.
\eq
When $\hetaH > -1/2$, both radial and azimuthal drift of the field exceeds corresponding drift due to fluid advection. In this case, large range of wavenumbers in region I (finite $k_c$ and $k_0$) in Fig.~(\ref{fig:FO}) are destabilised. 

To summarise, when only radial field drift $\hdvBr$ exceeds radial advection of the fluid $\hdvr$, magnetorotationally unstable waves of arbitrary $k_c$ and $k_0$ (corresponding to all three region I, II and III) in Fig.~(\ref{fig:FO}) is assisted by Hall diffusion. However, when both radial and azimuthal field drift exceed respective component of the fluid advection velocity, unstable waves with only finite $k_c$ and $k_0$ (corresponding to region I) is assisted by Hall diffusion.    

The maximum growth rate in the Hall—-Ohm diffusion plane can be easily inferred by  noting that the radius of the circle is determined by the non-zero shifted part of the $\eta_O$-axis in Fig~(\ref{fig:FO}). Thus setting $\eta_O = 0$ in Eq.~(\ref{eq:dno1}) and equating  the shifted part to the right hand side of Eq.~(\ref{eq:dno1}), i.e. $(\bs\,A)^2 = (3\,A / 4)^2$ yield
\bq
{\sigma}_{0} = \frac{3}{4}\,|\hk \cdot \Om|\,.
\label{eq:mgx}
\eq 
Note that above maximum MRI growth rate in diffusive Hall-—Ohm dominated disc is identical to the fully ionised, ideal MHD discs. This is not surprising since behaviour of the fluid in Hall--MHD is similar to the ideal MHD except that instead of magnetic flux freezing, generalized flux $\int{\left(\omega_H\,\hB + \curl\v\right) \cdot d{\bf S}}$ is frozen in the fluid \citep{PW06, PW08}. We note that maximum growth rate depends on the orientation of the wavevector which is also similar to the ideal MHD case \citep{BH92b}. 

As has been noted above, the $\eta_H-\eta_O$ plane can be divided into three distinct regions which traces three distinct form of $\sigma$ as a function of $k^2$. In region I, both most unstable wavenumber, $k_0$ and cut—-off wavenumber, $k_c$ are finite whereas in region II corresponding to $\hetaH \in [-2, -1 / 2]$, $k_0$ is finite and $k_c \rightarrow \infty$. In region III,  both $k_0$ and $k_c \rightarrow \infty$. Rescaling Eq.~(B11) of WS11, region I, in Fig.~(\ref{fig:FO}) outside the semi—-elliptical locus is described by  
\bq
\hetaO^2 + \mu^2\,\left[\hetaH + \frac{5}{4}\Bz\right]^2  = 
\frac{9}{16}\,\mu^2\,\Bz^2\,, 
\label{eq:sc1}
\eq
with eccentricity $\sqrt{1-–\mu^2} / \mu$. 

The maximum growth rate derived by rescaling Eq.~(B13) of WS11 yield   
\bq
\hetaH =  \frac{24\,\hs_0}{9\,\kz^2\,\Bz –- 16\,\hs_0^2\,\Bz}\,\hetaO - \frac{2\,\kz\,\Bz}{\kz^2 + \hs_0^2}\,,
\label{eq:OH1}
\eq 
and as shown in Fig.~6 by WS11, the contours of constant $\hs_0$ are straight lines in region I. The orientation of the field and wavevector merely modifies the slope of the curve.  

It is clear from Eq.~(\ref{eq:mgx}) that MRI growth rate is maximum when $\kz = 1$. We can also infer from Eq.~(\ref{eq:OH1}) that only vertical field threaded disc ($\Bz = 1$) grows at maximum rate. In order to see this, we plot Eq.~(\ref{eq:OH1}) for constant $\sigma_0$ with $\kz = 1$ in $\eta_O-\eta_H$ plane for $\Bz = 1$ (solid line) and $\Bz = 0.5$ (dashed line).  Making use of Eq.~(\ref{eq:dno1}), we also show corresponding contours for above values of field which subdivides  the diffusion plane between finite and infinite $k$. We see that when $\Bz = 1$, maximum growth rate line has a higher slope in $\eta_H-\eta_O$ plane than when $\Bz = 0.5$. This implies that MRI will grow at maximum rate only when the disc is threaded by purely vertical field. Therefore, MRI grows at maximum rate in discs having only vertical field and waves propagating along the field ($\kz = 1$).      
\begin{figure}
\includegraphics[scale=0.32]{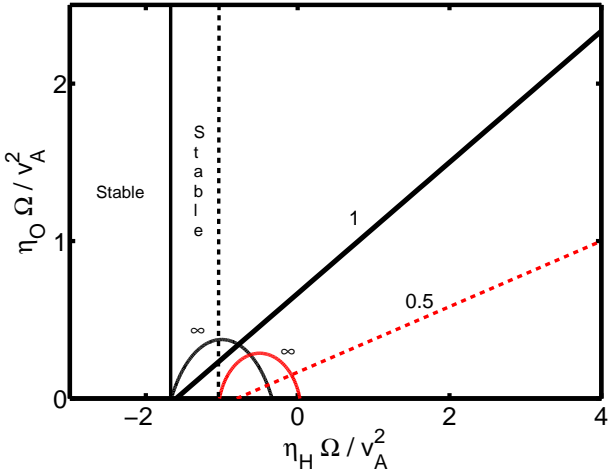}
\caption{The solution of Eq.~(\ref{eq:OH1}) for $\kz = 1$ and $\sigma_0 = 0.5\,\Omega$. Contours of $\sigma_0 = 0.5\,\Omega$ (straight lines) for $\Bz = 1$ (solid) and $\Bz = 0.5$ (dotted) is shown in the figure. The corresponding semi—circles Eq.~(\ref{eq:dno1}),  are the boundaries between regions I and II (like in Fig.~(\ref{fig:FO})) for $\Bz = 1$ (solid) and $\Bz = 0.5$ (dotted) and $k \rightarrow \infty$ is also shown in the figure.}
\label{fig:FO1}  
\end{figure}

The most unstable wavenumber becomes 
\bq
\left(\frac{k_{0}\,v_A}{\Omega}\right)^2 = \frac{-2\,\hs_0^2}{ \left(2\,\hs_0\,\hetaO -3/2\,s\,\hetaH\right) + \frac{\mu^2\,\left(- 3\,\kz^2 + 2\,\hs_0^2\right)}{\left(\kz^2 + \hs_0^2\right)}}\,. 
\label{eq:k01}
\eq
When vertical field is weak and Ohm diffusion is negligible, $k_0$ can become arbitrarily small. However, limit on the field is set by vertical scale height $\mbox{H} \sim c_s / \Omega$ of the disc owing to the requirement $k_0\,\mbox{H} \gtrsim 1$.     

The parameter window of the instability, is between interval $[0, k_c]$ where Eq.~(\ref{eq:kcc}) for Hall—-Ohm case becomes   
\bq
\left(\frac{k_{c}\,v_A}{\Omega}\right)^2 = \left(\frac{3\,s}{2}\right) \frac{\hetaH + 2\,\Bz}{\hetaO^2 + \mu^2\,\left(\hetaH + 2\,\Bz\right)\,\left(\hetaH + \frac{1}{2}\,\Bz\right) }\,. 
\label{eq:kc1}
\eq 

Region I encompasses both ideal-—MHD as well as diffusive limits. The inner boundary of region II occurs when $k_0 \rightarrow \infty$ and $\sigma_0$ is given by Eq.~(\ref{eq:OH1}). Following expression for the locus separating regions II and III is derived from WS11 [Eq.~(B15)]
\bq
\hetaH = - \frac{2\,\kz^2\,\Bz\,\left(9\,\kz^2 + 4\,\hs_0^2\right)}
{\left(\kz^2 + \hs_0^2\right) \left(9\,\kz^2 + 16\,\hs_0^2\right)}\,,
\eq
\bq
\hetaO = \frac{\kz^2\,\Bz^2\,\hs_0\left(9\,\kz^2 -– 16\,\hs_0^2\right)}
{\left(\kz^2 + \hs_0^2\right) \left(9\,\kz^2 + 16\,\hs_0^2\right)}\,,
\eq
which traces an arc in $\hetaH-\hetaO$ plane from $\hetaO-\hetaH = (0, - 2\,\Bz)$ to $\hetaO-\hetaH = (0, - \frac{4}{5}\,\Bz)$ as $\hs_0$ runs from $0 $ to $0.75\,\kz$. Therefore, with decreasing vertical field, region III becomes smaller (Fig.~\ref{fig:FO}). 

We conclude from above analysis that Hall—-Ohm diffusion dominated discs with very simple field geometry, having only vertical magnetic field and vertical wavevector plays fundamental role in destabilising it against axisymmetric fluctuations. Since MRI grows at maximum rate in such a disc, we may infer that $\B = (0, 0, B_z)$ and $\k = (0, 0, k_z)$ is the most magnetorotationally favoured topology in the disc. The role of the azimuthal field and radial wavevector is limited to scaling various regions of instability as shown in Fig.~(\ref{fig:FO}).       

\subsection{Hall—-Ambipolar diffusion plane}
The ambipolar diffusion is the dominant diffusive mechanism in the surface layer of the protoplanetary disc followed by the Hall in the middle layers. This generic picture is broadly valid close to the protosolar nebula as well as further out in the disc at $10\,\mbox{AU}$ even when grains have not settled to the midplane of the disc \citep{W07, B11}. Therefore, it is important to analyse the dispersion relation in the Hall—-Ambipolar diffusion dominated disc.       

By setting $\eta_O = 0$ the dispersion relation, Eq.~(\ref{eq:dnc}) can be written as 
\bq
\eta_H'^2 + \eta_A'^2  = N\,\left(\frac{1}{1 + \bs^2} + \frac{1}{\bk^2}\right)^2\,,
\label{eq:dnom}
\eq
where 
%\begin{eqnarray} 
\begin{eqnarray}
\eta_H' = \betaH + \frac{5/4}{1 + \bs^2} - \frac{3/4}{\bk^2}\,,\nonumber\\
\eta_A' =\mu\,\left[\betaA + \frac{1}{2\,\mu^2}\left(\frac{1}{1 + \bs^2} + \frac{1}{\bk^2}\right) \left[\left(1 + \mu^2\right)\bs - \frac{3}{2}\,g^{\prime}\right]\right]\,,
\end{eqnarray}
%\nonumber\\
and
%\nonumber\\
\begin{eqnarray} 
N = \frac{1}{\mu^2}\,\left[ \frac{9}{16} (\mu^2 + g{^{\prime}}^2) + \frac{\bs^2\,\left(1 - \mu^2\right)^2}{4} - \frac{3}{4} \bs\,g^{\prime}\,\left(1+ \mu^2\right)\right]\,.
\label{eq:ams}
\end{eqnarray}
This dispersion relation describes a circular locus in $\eta_A-\eta_H$ plane for given $\sigma$ and $k$. 

In order to find the maximum growth rate in ambipolar diffusion dominated disc, we set $\eta_O = \eta_H = 0$ in Eq.~(\ref{eq:mdr}) and treat it as an equation in $\oa^2 \equiv \mu^2\,k^2\,v_A^2$, and $k^2$ in $\eta$ plane. Taking the partial derivative in $\oa^2$, and $k^2$ and eliminating $\oa^2$ from one of the equations, we arrive at the following equation for the maximum growth rate
\begin{eqnarray}
\betaA\,\left\{
\left(1 - \mu^2\right)\,\bk^2\,\betaA\,\bs_0^2  - \left(1 + \mu^2\right)\left(5 + 3\,g^{\prime}\,\bk^2\,\betaA\right)\,\,\bs_0 \right.\nonumber\\ \left.
- 0.25\left(16\,\mu^2 -– 9\,g^{\prime ^2}\right)\,\bk^2\,\betaA + 
\frac{15}{2}\,g^{\prime} \right\} = 0\,. 
\label{eq:amx1}
\end{eqnarray}
Although a general expression for the maximum growth rate can be easily written from above equation, it acquires particularly simple form in the limiting cases. In the long wavelength limit ($k \rightarrow 0$) the maximum growth rate becomes 
\bq
\sigma_0 \approx \frac{3\,g\,\Omega}{2\,\left(1 + \mu^2\right)}\,.
\label{eq:amx3}
\eq
It is not surprising that the maximum growth rate in ambipolar diffusion dominated disc is proportional to $g$ which encapsulates the topology of the disc field and obliqueness of the wavevector \citep{D04}. The parameter $g$ couples the poloidal wavevector to the toroidal magnetic field. Since the rotational velocity $v_{\phi} = r\,\Omega$ is also along the azimuthal direction, the presence of non–zero azimuthal field along $v_{\phi}$ allows the fluctuations to tap into the reservoir of free energy. Therefore, topological factor $g$ acts like a switch to the reservoir and if this factor is positive, then anisotropic ambipolar diffusion is able to assist the instability.  

Recall that the stability condition, Eq.~(\ref{eq:pwd}) which is long wavelength limit of the sufficient condition, $C_0 < 0$ requires positive $g$. As $g \in [-0.25, 0.25]$, we conclude that the maximum growth rate in purely ambipolar regime is less than half of the ideal MHD or, Hall-—MHD regime for $\mu = 1 / 2$.  
Since anisotropic dissipation is responsible for MRI in ambipolar diffusion dominated discs \citep{D04}, part of free energy, otherwise available to waves in non—dissipative discs is lost to dissipation causing smaller MRI growth rate. 

The maximum growth rate in short wavelength ($k \rightarrow \infty$) becomes
\bq
\sigma_0 \approx \frac{3\,\left(1 + \mu^2\right)g\,\Omega}{2\,\left(1 - \mu^2\right)} \left(1 - \frac{\mu\,\sqrt{\Delta}}{3\,\left(1 + \mu^2\right)g}\right)
\label{eq:amxx}
\eq
 where $\Delta =  27\,g^2 + 9\,\mu^2\,g^2 + 16\,\kz^2\,\left(1 - \mu^2\right)$. Dropping terms of the order of $\left(\gp\right)^2$, $\sigma_0$ can be written as
\bq
\sigma_0 \approx \frac{3\,\left(1 + \mu^2\right)g\,\Omega}{2\,\left(1 - \mu^2\right)} - \frac{2\,s}{\sqrt{\left(1 - \mu^2\right)}}\,
\label{eq:amxx}
\eq
which shows that short wavelength fluctuations are largely damped. For example when $g = 0.25$ and $\kz = \Bz = 1 / \sqrt{2}$, $\sigma_0 = - 0.19\,\Omega$.

Since growth rate in the ambipolar regime is proportional to $g$ we may conclude that predominantly Alfv\'enic ($\mu \rightarrow 1$) or, predominantly magnetosonic ($\mu \rightarrow 1$) fluctuations will grow at very small rate since $g \rightarrow 0$. However, oblique waves with $\kz = \kr = 1 / \sqrt{2}$ and $\Bz = - \Bp = 1 / \sqrt{2}$ grows at maximum rate. Therefore, oblique magnetic field inclined at $\pi /4$ in $\phi – z$ plane is most favoured MRI geometry in ambipolar diffusion dominated discs. 
         
\begin{figure}
     \includegraphics[scale=0.24]{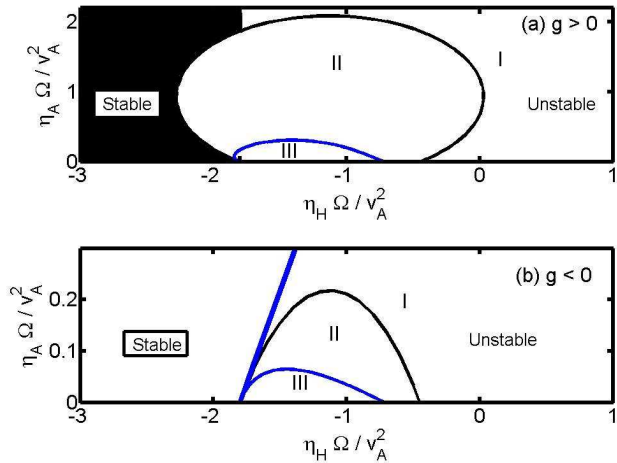}
     \caption{The solution of Eq.~(\ref{eq:dnom}) for $\kz = 0.3$, $\Bz = 0.9$  are shown in the top [$g > 0$ ($\Bp < 0$)] and bottom [$g < 0$ ($\Bp > 0$)] panels respectively. The vertical line separates stable and unstable regions. The unstable region is subdivided in three different regions corresponding to different $k_0$ and $k_c$ as shown in Fig.~(\ref{fig:FC}).}
 \label{fig:abr}  
\end{figure}

In Fig.~(\ref{fig:abr}) the locus of circle for both positive and negative $g$ is shown. The vertical line separates the stable region from the unstable region. The sign of $g$ is changed by changing the sign of azimuthal magnetic field. It is clear from Fig.~\ref{fig:abr}(a)--(b) that when $g > 0$, area under the curve in $\eta_H-\eta_A$ plane is larger than when  $g < 0$. This happens because ambipolar diffusion is conducive to MRI for positive $g$. Note that on the horizontal Hall axis, if $\hetaH > -2\,\Bz -– g\,\hetaA / s$, the Hall diffusion will assist MRI. Therefore, for positive $g$ and $\eta_A \neq 0$, Hall diffusion can affect MRI over regions which were otherwise inaccessible in a purely vertical field threaded disc (Fig.~5, WS11 or, Fig.~1-—3, BT01). Therefore, when $g > 0$, ambipolar diffusion not only assists MRI but simultaneously allows Hall diffusion to operate in enlarged parameter space. As a result the locus dividing region I and II in Fig.~\ref{fig:abr}(a) has an outward bulge. Similarly, negative $g$, $\eta_A \neq 0$ causes shrinkage of parameter space for Hall diffusion [Fig.~\ref{fig:abr}(b)]. Clearly, favourable field topology and obliqueness of wavevector not only affects MRI in ambipolar diffusion dominated disc but also extend the domain over which Hall can operate in the disc. Therefore, the sign of $g$ plays crucial role in affecting the stability of Hall-—ambipolar dominated discs.     

\begin{figure}
     \includegraphics[scale=0.24]{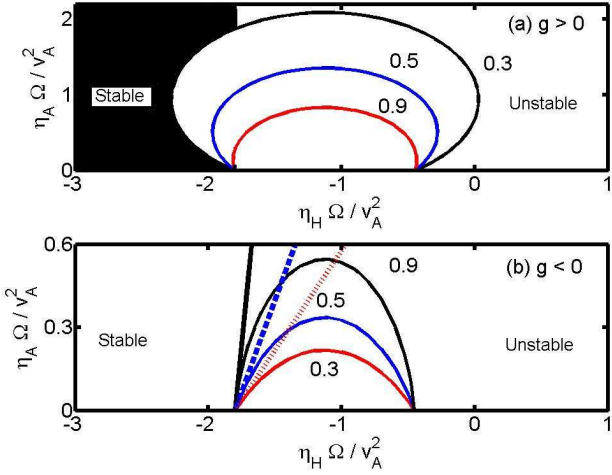}
     \caption{The dependence of locus on $\kz$ diving region I and II [ Fig.~(\ref{fig:abr})] for $\Bz = 0.9$  are shown in the top [$g > 0$ ($\Bp < 0$)] and bottom [$g < 0$ ($\Bp > 0$)] panels respectively.}
 \label{fig:Noh1}  
\end{figure}
In Fig.~\ref{fig:Noh1} we plot locus of the circle dividing regions I and II shown in Fig.~(\ref{fig:abr}) for various $\kz$. We see that with decreasing $\kz$, area under the curve increases. This implies that almost radial fluctuations of arbitrary wavelength is unstable in AD dominated disc, in conformity with the general criterion Eq.~(\ref{eq:pwd1}) or, Eq.~(\ref{eq:pwd2}). Although accessible $\eta_A-\eta_H$ parameter window increases with decreasing $\kz$, the growth rate of the instability, which is proportional to $g \propto \kz$ becomes small. When $g < 0$, radius of the semi-circle locus decreases with decreasing $k_z$, implying that almost radial fluctuations are severely damped. 

Like Hall—-Ohm case, Hall--ambipolar plane can also be divided into three distinct regions. In region I, outside the semi—-circular locus  
\bq
\left(\betaH + \frac{5}{4}\right)^2 + \mu^2\,\left(\betaA - \frac{3}{4\,\mu^2}\,\gp\right)^2 = 
\frac{9}{16\,\mu^2}\,\left(\mu^2 + {g^{\prime}}^2\right)\,,
\label{eq:sc1}
\eq
describes a half-—circle in the $\eta_H-\eta_A$ plane. Region I encompasses both ideal MHD and ambipolar—-Hall space. All the wavenumbers within region II inside the semicircle, described by Eq.~(\ref{eq:sc1}) are unstable.

The maximum growth rate in $\eta_A - \eta_H$ plane can be found by setting the discriminant of Eq.~(\ref{eq:dnc}) to zero. This gives
\bq
\betaH = \frac{\Bz\,R_1 / \kz}{1 + \bs_0^2}\,\hetaA - \frac{2}{1 + \bs_0^2}\,,
\label{eq:mdh}
\eq
Here
\bq
R_1 = - \frac{S \sqrt{\frac{-q_2}{q_1}} + T}{S - \sqrt{\frac{-q_2}{q_1}}\,T}\,, 
\eq
\bq
S = \frac{b_1}{\sqrt{b_1^2 + \left(q_1 –- a_1\right)^2}}\,\,,
T = - \frac{S\,\left(q_1 -– a_1\right)}{b_1}\,,
\eq
\bq
q_{1,2} = \frac{1}{2}\,\left[\left(a_1 + c_1\right) 
\mp \sqrt{\left(a_1 -– c_1\right)^2 + 4\,b_1^2} \right]\,,
\eq
and
\begin{eqnarray}
a_1 = \Bz^2\,\left(9 -– 16\,\bs_0^2\right)\,,\quad
b_1 = 6\,\mu\, \left(1 + \bs_0^2\right)\,d_0\,,\nonumber\\
c_1 = 4\,\kz^2\,\left(1 + \bs_0^2\right)^2\,
\left(d_0^2 -– 4\,\mu^2\,\bs_0^2\right)\,,
\end{eqnarray}   
with $d_0 = \left(1 + \mu^2 \right)\,\bs_0 - \frac{3}{2}\,g^{\prime}$. If $B = B_z$ and wave is propagating along the field, then Eq.~(\ref{eq:mdh}) reduces to Eq.~(B13) of WS11 since     
\bq
R_1 = \frac{24\,\hs_0\,\left(1 + \hs_0^2\right)}
{\left(9 –- 16\,\hs_0^2\right)}\,.
\eq
The wavenumber corresponding to the maximum growth rate is $-b / 2\,a$ which is found by setting the discriminant $b^2 -– 4 a c = 0$ in Eq.~(\ref{eq:dnc}).

In region I, both $k_0$ and $k_c$ are finite whereas in region II, $k_0$ is finite and $k_c \rightarrow \infty$. Region III in Fig.~(\ref{fig:abr}) is approached asymptotically when both most unstable wavelength, and critical wavelength approach infinity. The dispersion relation (\ref{eq:dnom}) in $k\,v_A / \Omega \rightarrow \infty$ limit becomes
\begin{eqnarray}
\left(\betaH + \frac{5 / 4}{1 + \bs_0^2}\right)^2 + 
\mu^2\,\left(\betaA + \frac{d_0}{2\,\mu^2\,\left(1 + \bs_0^2\right)}\right)^2 
\nonumber\\
= \frac{N}{\left(1 + \bs_0^2\right)^2}\,.
\label{eq:mdh1}
\end{eqnarray}         
The locus that separates region II and III, can be found by equating Eqs.~(\ref{eq:mdh}) and (\ref{eq:mdh1}). This gives the following parametric solution 
\bq
\betaH = \frac{4\,\mu\,R_1\,\bs_0^2\,/ \left(1 + \bs_0^2\right)}{3\,\mu\,R_1 –- 2\,\kz^2\,\left(1 + \bs_0^2\right)\,d_0}  
- \frac{2}{1 + \sigma_0^2}\,
\label{eq:mdh2}
\eq
\bq
\betaA = \frac{4\,\kz\,\bs_0^2}{3\,\Bz\,R_1 –- 2\,\kz\,\left(1 + \bs_0^2\right)\,d_0}\,,
\label{eq:mdh3}
\eq
which traces an arc from $(-2\,\Bz, 0)$ to $(\eta_H, \eta_A)|_{\sigma_0})$ in the diffusive plane.

We see from Eq.~(\ref{eq:kcc}) that the 
wavenumber below the critical value
\bq
\left(\frac{k_{c}\,v_A}{\Omega}\right)^2  = 
\frac{\frac{3}{2}\left[s\,\hetaH + 2\,\mu^2 + g\,\hetaA\right] / \mu^2}
{
\hetaA^2 + \left(\hetaH + 2\,\Bz\right)\, \left(\hetaH + \frac{1}{2}\,\Bz\right)
 - \frac{3\,g}{2\,\kz^2}\,\hetaA} \,,
\label{eq:kc3}
\eq
are unstable in the Hall—-ambipolar disc. Only when the wavevector is aligned to the ambient field, i.e. $\mu = 1$, Eq.~(\ref{eq:kc3}) becomes identical to Eq.~(\ref{eq:kc1}) if we replace $\eta_A$ with $\eta_O$. However, even for a purely vertical field, Eq.~(\ref{eq:kc3}) does not reduce to Eq.~(\ref{eq:kc1}). Note that when wavevector is aligned to the ambient field, i.e. $\mu = 1$, Ohm and ambipolar diffusion can be combined together as Pedersen diffusion and their effect on the wave is identical--they both cause damping (WS11). However, when $\mu \neq 1$, whereas Ohm diffusion isotropically damps the waves, damping by ambipolar diffusion is anisotropic. Therefore, $\mu \neq 1$, lifts the diffusive degeneracy causing ambipolar diffusion to assist MRI. Thus, critical wavelength in this case can not be reduced to Hall—-Ohm case even for a purely vertical field.   

To conclude, anisotropic nature of ambipolar diffusion which is encapsulated in the topological factor $g$, plays crucial role in assisting MRI in Hall—-ambipolar diffusion dominated discs. Not only positive $g$ assists MRI in ambipolar regime but simultaneously enlarges the parameter space over which Hall can influence MRI. In a purely ambipolar diffusion dominated disc, maximum growth rate is proportional to $3\,g / 4$ when the wave is almost vertical whereas it is $3\,g / 2$ for almost radial waves.

\section{Limiting cases of the dispersion relation}
In the {\it standard} MRI picture pertaining to fully ionised, differentially rotating disc radial field is generated by the inward motion of the fluid element and the rotational shear generates azimuthal magnetic field ($\dBp$) from radial field ($\dBr$) \citep{BH91}. In a partially ionised, diffusive disc, not only differential rotation but both Hall and anisotropic ambipolar diffusion are also capable of generating azimuthal field from the radial field. Therefore, magnetic diffusion provides additional new pathways through which free shear energy can be transferred to waves. Even when one of these pathways (through which free energy is channelled to waves) is blocked, disc can still become magnetorotationally unstable in the presence of diffusion. To see this, we first note that three terms in Eq.~(\ref{eq:lin2}) are $\sim 1\,,\hk^2\,,\hk^2\,\eta$ and thus, one of the following three scenarios may exist in a diffusive disc (WS11).

A. {\it Ideal MHD:\footnote{The term ideal MHD is misnomer here since disc matter is partially or, weakly ionised.} Diffusion in the disc is weak in comparison with the fluid advection, i.e. field is frozen in the fluid; $\left[\hdv  \sim \hdB\right]$}.  In this limit $\hk^2 \sim 1 \gg \hk^2\,\heta$, i.e. $\heta \ll 1$ and last term in Eq.~(\ref{eq:lin2}) can be neglected. As noted above, presence of rotational shear in the disc generates $\dBp$ from $\dBr$ in this case.   

B. {\it Cyclotron limit: Diffusion in the disc is comparable to fluid advection:} $\left[\hdv  \gg \hdB\right]$.  
In this case $\hk^2\,\heta \sim \hk^2 \gg 1$. This is the low frequency limit since $\hs \sim 1$ and first term in Eq.~(\ref{eq:lin2}), can be neglected. The low frequency, short wavelength dressed ion—-cyclotron wave with frequency $\omega_{C} = \mu\, \omega_H$ \footnote{The cut—-off frequency $\omega_{C} = \omega_H$, Eq.~(41) given by \cite{K08} correspond to $\mu =1$ only.} is the normal mode of the system \citep{PW08}. Therefore, when wave is propagating along the vertical field ($\mu = 1$), the cut—-off frequency $\omega_C$ is solely determined by the fractional ionization $\omega_H / \omega_{ci} \sim X_e$.  In PPDs where fractional ionization is small \citep{WN99, W07, B11}, left circularly polarized waves will have extremely low cut-—off frequency. However, the role of fractional ionization in determining the cut-—off becomes redundant if wave in the medium is propagating almost radially ($\mu \rightarrow 0$). In this case, waves will have very low cut--off frequency. Therefore, not only the fractional ionization, but also the direction of the wavevector determines the cut--off threshold of left circularly polarised ion—-cyclotron waves.

The MRI in the cyclotron limit also requires like {\it standard} MRI picture (a) the generation of radial field due to inward fluid motion and (b) the generation of $\dBp$ from $\dBr$. However, unlike standard MRI, the azimuthal field is not generated by the differential shear of the radial field but by the Hall, or, anisotropic (reflected through topological switch $g$) ambipolar diffusion.  Clearly, step (b) of the standard MRI is now replaced by a totally new step available only to diffusive discs. To distinguish this unique feature of MRI in diffusive discs, we shall call this case diffusive MRI or, DMRI.    

\begin{figure}
     \includegraphics[scale=0.32]{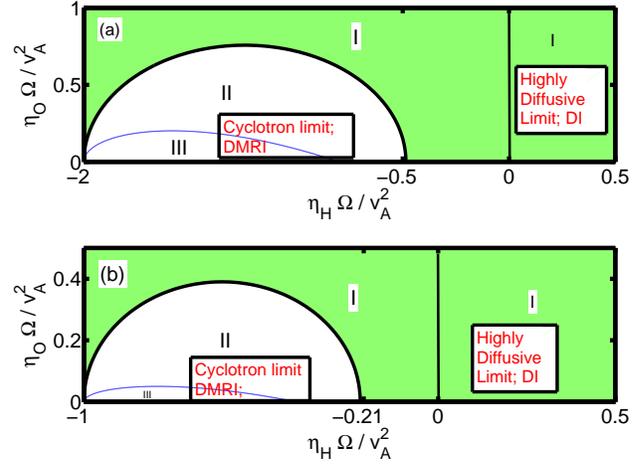}
     \caption{The Diffusive MRI (cyclotron limit, regions II and III) and Diffusive Instability (highly diffusive limit, region I, $\eta_H > 0$) of Fig.~(\ref{fig:FO}) are highlighted in the above figure.}
 \label{fig:lfx}  
\end{figure}

Taking $k \rightarrow \infty$, and thus setting $a(\sigma) = 0$ in Eq.~(\ref{eq:dnc}) or, neglecting the left hand side of Eq.~(\ref{eq:lin2}), we get following dispersion relation
\begin{eqnarray}
\left(\betaH^2 + \betaP\,\betaT\right)\,\bs^2 + \left(\betaP + \betaT\right)\,\bs + 2\,\left(2 - q\right) \,\left[ \betaP\,\betaT
\right. 
\nonumber\\
 \left.
+ \left(\betaH + \frac{1}{2 - q}\right)^2 \right]
%\nonumber\\
= q\,\left(g^{\prime}\,\betaA + \betaH + \frac{1}{2 - q}\right)\,.
\label{eq:EsL} 
\end{eqnarray}
Since coefficients on the left hand side of preceding equation is positive, a positive $\sigma$ requires that the right hand side be positive. Thus we arrive at the stability criterion which is identical to Eq.~(\ref{eq:pwd}). This is not surprising since DMRI is a subset of MRI and corresponds to region II and III in Fig.~(\ref{fig:lfx}). We have used here Hall—-Ohm diffusion dominated disc to highlight cyclotron and highly diffusive limits. However, Hall—-ambipolar dominated disc, e.g. Fig.~(\ref{fig:abr}) can as well be used for the same pupose.

The regions II and III in Fig.~(\ref{fig:lfx}) represent cyclotron limit can be directly inferred from marginal stability condition [$C_0 = 0$] also. Thus from Eq.~(\ref{eq:coefm}) along the Hall axis we get
\bq
\hetaH = - \frac{1}{2} + \frac{q}{\hk^2}\,,\quad
\hetaH = - \frac{1}{2 - q}\,.
\eq 
For a Keplerian disc ($q = 3 / 2$) in $k \rightarrow \infty$ limit, above equations describes the interval between $\hetaH = [-2\,, -1 / 2]$ in Figs.~(\ref{fig:FO}) and (\ref{fig:abr}). 

Assuming $\kz < 1$ and thus neglecting $\kz^2$ terms, for a Hall dominated Keplerian disc, from Eq.~(\ref{eq:EsL}) we get     
\bq
\sigma \approx \left[\frac{5}{2}\,s\,\Omega\omega_H\right]^{1/2} = \left[\frac{5}{2}\,\kz\,\Omega\,\omega_{C}  \right]^{1/2}\,.
\label{eq:ic1}
\eq
Thus the growth rate of ion—-cyclotron wave is about one and half times the geometric mean of the rotation and ion—-cyclotron frequencies and attains maximum value only when $\kz = 1$.   

In a purely ambipolar diffusion dominated disc, i.e. setting $\eta_O = \eta_H = 0$, and assuming $\mu \rightarrow 0$, i.e. when wavenumbers are almost radial, we get
\bq
\sigma \approx \frac{3}{2}\,g\,\Omega\,,
\label{eq:ahfmax1}
\eq 
which is identical to the growth rate, Eq.~(\ref{eq:amx6}). Hence, both DMRI and MRI grow at similar rate in ambipolar diffusion dominated disc. 

C. {\it Highly diffusive limit: Magnetic Diffusion in the disc overwhelms fluid advection;}$ \left[ \hdv  \ll \hdB \right]$. 
In this case $\hk^2\,\heta \sim 1 \gg \hk^2$, i.e. $\heta \gg 1$ and $\hk^2 \ll 1$. The evolution of magnetic field which is kinematic in nature in this case is solely due etermined by the diffusion. Only first and last term in Eq.~(\ref{eq:lin2}) are retained.  

In the presence of rotational shear, diffusion of magnetic field alone is sufficient to cause the instability in the disc \citep{RK05}. It should be pointed out that rotation is not essential for this instability. The presence of plane shear flow can as well cause this instability. As has been noted by \cite{K08}, this is a new kind of instability –- diffusion instability (DI). 

Unlike MRI or, DMRI, the angular momentum transport in the disc by DI is not efficient. The physical mechanism of DI is quite simple: rotational shear generates $\dBp$ from $\dBr$ and diffusion completes the loop by generating $\dBr$ from $\dBp$, i.e.
\bq
\dBr \xrightarrow[]{{\bmath{\Omega}}\left(r\right)} \dBp\,,\quad
\dBp \xrightarrow[]{{\bmath{\eta}}} \dBr\,.
\eq
It is important to note that although both Hall and ambipolar diffusion generate $\dBr$ from $\dBp$ the generation mechanism in two cases are quite different. Whereas, Hall effect through dissipationless diffusion generates $\dBr$ from $\dBp$, the ambipolar diffusion does it via anisotropic dissipation \citep{D04}. Therefore, suitable disc field and normal wavevector is necessary condition for ambipolar diffusion to act like Hall. For purely vertical field, when $g = 0$, ambipolar diffusion is incapable of generating $\dBr$ from $\dBp$. In this case, like Ohm it can only cause the damping of waves.  

The dispersion relation for a Keplerian disc in the highly diffusive limit becomes
\bq
\hs^2 + \hk^2\left[2\,\hetaO + \left(1 + \mu^2\right)\,\hetaA\right]\,\hs + D = 0\,,  
\label{eq:HDL1}
\eq
where
 \begin{eqnarray}
D = \hk^4\,\left[\hetaO^2 + \left(1 + \mu^2\right)\,\hetaO\,\hetaA + \left(1 -– \kr^2\,\Bz^2\right)\,\Bz^2\,\hetaA^2 \right] \nonumber\\
- \hk^2 \left[\frac{3}{2}\left(g\,\hetaA + s\,\hetaH\right) + \frac{\hk^2}{\kz^2}\left(g^2\,\hetaA^2 -– s^2\,\hetaH^2\right)\right]\,.
\end{eqnarray}
In the absence of $\hetaH$, above dispersion relation reduces to Eq.~(39) of \cite{D04}. 

Eq.~(\ref{eq:HDL1}) can be recast in the following form 
\begin{eqnarray}
\left(\hetaP\,\hetaT + \mu^2\,\hetaH^2\right)\,\hk^4 + \left(\hetaP + \hetaT\right)\,\hs\,\hk^2 + \hs^2 
\nonumber\\
= \frac{3}{2}\,\hk^2\,\left(g\,\hetaA + s\,\hetaH\right)\,, 
\end{eqnarray}
from where it is easy to infer general stability criterion for a diffusive disc 
\bq
g\,\hetaA + s\,\hetaH > 0\,.
\label{eq:gdi}
\eq 
Clearly excessively diffusive Keplerian disc is unstable if the linear combination of ambipolar and Hall diffusion with suitable topological coefficients is positive.  In the Hall—-Ohm dominated regime if $s\,\eta_H > 0$ the disc becomes diffusively unstable.  A comparison of Eq.~(\ref{eq:gdi}) with Eq.~(\ref{eq:pwd}) shows that if the disc is unstable due to DI then it is also unstable due to DMRI. However, reverse is not true.  This can also be seen from Figs.~(\ref{fig:lfx}). Since DI belongs to region I [beyond $\eta_H = 0$ along the Hall axis, in Fig.~(\ref{fig:lfx}) corresponding to finite $k_0\,,k_c$, Fig.~(\ref{fig:FC})] guaranteeing that Eq.~(\ref{eq:pwd}) is satisfied in the Hall—-Ohm dominated discs. 

In the absence of ambipolar diffusion, as has been shown in section 2.2, the dispersion relation, Eq.~(\ref{eq:HDL1}) for Hall—-Ohm dominated disc can be mapped to a purely vertical field threaded disc which has been analysed in considerable detail by WS11. Therefore, we shall focus only on Hall—-ambipolar diffusion dominated disc here.   

The instability in Hall-—ambipolar diffusion dominated disc will grow at maximum rate
\begin{eqnarray}
\sigma_0 = \frac{3\,\Omega}{2}\frac{\left(g\,\hetaA + s\,\hetaH\right)}{\left[ 
\left(1 - \mu^2\right)\,\hetaA^2 -– 4\,\mu^2\,\hetaH^2\right]}
\left[\left(1 + \mu^2\right)\,\hetaA\right.\nonumber\\
\left.+ 2\,\mu\,\sqrt{\hetaA^2 + \hetaH^2}
\right]\,,
\label{eq:hamx}
\end{eqnarray}
which for purely Hall case ($\eta_A = 0$) reduces to Eq.~(\ref{eq:mgx}). For purely ambipolar ($\eta_H = 0$) case, preceding equation becomes
\bq
\sigma_0 = \frac{3}{2} \frac{g\,\Omega}{\left(1 + |\mu| \right)^2}\,,
\label{eq:ahfmax}
\eq
which is similar to Eq.~(\ref{eq:amx3}). For planar shear gradient ($d v / dx = 3\,\Omega/2$) above equation becomes identical to Eq.~(33) of \cite{K08}.  
Note that the maximum growth rate, Eq.~(\ref{eq:ahfmax})  is proportional to the shear gradient and is independent of $\eta_A$ as well as the strength of the background magnetic field. However, signature of the magnetic field is manifested through the geometric factors $g$ and $\mu$. 

The corresponding most unstable wavevector is also a function of $g$ since
\bq
\left(\frac{k_0\,v_A}{\Omega}\right) = \sqrt{\frac{3\,g}{2\,\mu\,\left(1 + \mu \right)^2\,\hetaA}}\,.
\eq
Apart from the shear gradient and geometric factor, the wavevector $k_0$ also depends on the strength of the ambient magnetic field as $k_0 \propto 1/ B$. Thus in the presence of weak magnetic field extremely short wavelength fluctuations will grow at maximum rate in the ambipolar diffusion dominated discs.   

Notwithstanding the independence of maximum growth rate on diffusivities in a purely Hall or, purely ambipolar case, the general expression Eq.~(\ref{eq:hamx}), does depend on diffusivities. In fact, the maximum growth rate, decreases with increasing $\eta_A / \eta_H$. 
\begin{figure}
     \includegraphics[scale=0.32]{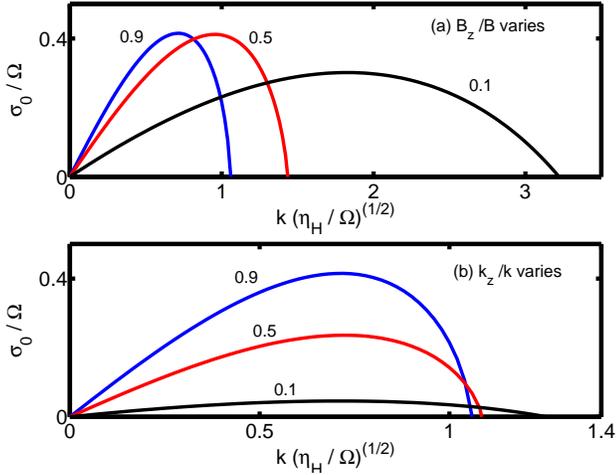}
 \caption{Maximum growth rate versus wavevector in the highly diffusive limit with fixed $\kz = 0.9$ and varying $\Bz$ is shown in 4(a) and fixed $\Bz = 0.9$ and varying $\kz$ is shown in 4(b). To keep $g$ positive, we have assumed $\Bp = - \sqrt{1 - \Bz^2}$.}
 \label{fig:F6HDL}  
\end{figure}

In order to see this dependence on the field geometry and wavevector, we plot in Fig.~[\ref{fig:F6HDL}(a)-—(b)] maximum growth rate for two different cases: (a) vary $\Bz$ and keep $\kz$ fixed and (b) vary $\kz$ and keep $\Bz$ fixed. We have assumed $\eta_A / \eta_H = 0.1$. The first panel Fig.~[\ref{fig:F6HDL}(a)] for fixed $\kz = 0.9$ suggests that when field is almost vertical, the diffusive instability grows almost at maximum rate. However, only long wavelength fluctuations, i.e. $ 2\,k^2\eta_H/3 \lesssim 1$ can be excited in this case. On the other hand in azimuthal field dominated disc, fluctuations of smaller wavelength also becomes unstable, albeit with reduced growth rate. We see from Fig.~[\ref{fig:F6HDL}(b)] that the instability will be completely quenched for small $\kz$. Therefore, in a predominantly azimuthal field threaded disc, almost vertical axisymmetric fluctuations are most prone to diffusive instability. 

The wave number between $[0, k_c]$ where
\bq
\left(\frac{k_c\,v_A}{\Omega}\right)^2 = \left(\frac{3}{2\,\mu^2}\right)\,
\frac{g\,\hetaA + s\,\hetaH}{\hetaA^2 + \hetaH^2}\,.
\eq 
are unstable in the highly diffusive disc. Note that when $\eta_H = 0$, the cut—off wavelength in AD dominated disc shifts to infinity when $g < 0$. As a result, AD damps fluctuations at all wavelengths. However, when $g > 0$, only fluctuations between $[0, k_c]$ can grow, where 
\bq
\left(\frac{k_c\,v_A}{\Omega}\right) = \sqrt{\left(\frac{3}{2}\right)\,
\frac{g}{\mu^2\,\hetaA}}\,.
\eq   
Choosing maximal value for $g = 1 / 4$, i.e. $\Bz = - \Bp =  1 / \sqrt{2}$, and $\kz = \kr = 1 / \sqrt{2}$, i.e. $\mu^2 / g = 1 $, we see from the above equation that waves of wavelengths $[0, 5\,\sqrt{\hetaA}]$ will be unstable in the AD dominated discs.    

To conclude, excessively diffusive discs can become unstable solely due to interplay between rotational shear and magnetic diffusion without significant transport of angular momentum. However, the onset of diffusive instability will always trigger MRI and thus transport of angular momentum will always occur in such discs. In discs threaded only by a vertical field both DI and DMRI grow at maximum rate when the wavevector has only vertical component. However, critical wavenumber window increases with decreasing vertical field.   
\section{Discussion}
Analysis of the MRI in three dimensional diffusion space for sufficiently general geometry is difficult. Often, protoplanetary discs (PPDs) have layered structure with surface layer well coupled to the magnetic field whereas midplane is magnetically dead \citep{G96}. Further, Hall—-Ohm is the dominant diffusion at $1—-2$ scale heights whereas ambipolar—-Hall dominates at $3—-4$ scale heights from the disc midplane \citep{W07, B11}. This provides a physical motivation to analyse a simplified dispersion relation in two dimensional Hall-—Ohm or Hall-—ambipolar diffusion plane which also lends to an easy geometric interpretation. Such a representation of the dispersion relation allows us to geometrically infer the maximum growth rate of the instability. However, the stability criterion Eq.~(\ref{eq:pwd1}) pertains to general diffusive disc and allows us to infer following conclusion.         

\subsection{The diffusive discs are likely to be unstable to almost radial fluctuations.} This can be seen from Eq.~(\ref{eq:pwd1}) which  can be written as 
\bq
\left(- \kr\,\Bp \right) > - \kz \frac{\left(\eta_H + 2\,\Bz\,v_A^2 / \Omega\right)}{\eta_A}\,.
\label{eq:pwd2}
\eq       
Assuming positive left hand side in the preceding equation above criterion is easily satisfied for non—-zero $\eta_A$ and $\kz \rightarrow 0$. Therefore, in the presence of almost radial fluctuations, the diffusive disc is always unstable provided modes with short enough wavelength fits in the disk, i.e. $k\,H \gtrsim 1$. 

Note that magnetic tension suppresses MRI growth for wavelengths $k^{-1} \lesssim v_A / \Omega \equiv H / \sqrt{\beta}$ (where plasma $\beta = c_s^2 / v_A^2$ is the ratio of magnetic to gas pressure). Only when the disc is subthermal ($\beta > 1$), MRI unstable fluctuations will fit in the disc. For example, in PPDs, it is quite possible that MRI active layer may reside in the upper disc when plasma $\beta$ is typically much larger than 1 \citep{B11}. Therefore if the upper layer in the inner PPDs ($\lesssim 10\,\mbox{AU}$) is subthermal, it may become magnetorotationally unstable to almost radial fluctuations.  At outer disc, even the disc midplane may become unstable to such fluctuations. However, non-subthermal diffusive discs, MRI modes can not fit within the disc scale heights. The disc wind may efficiently remove angular momentum in such diffusive discs \citep{WK93}. 

\subsection{Upper and middle layer of PPD is most susceptible to MRI and DMRI.}

As has been noted above, owing to its layered structure, diffusivities in PPDs vary with height and thus the critical wavelength which is a function of diffusivity also varies with height. In the upper layer of the disc when Hall—-ambipolar is the dominant diffusion, fluctuations of arbitrarily small wavelength becomes magnetorotationally unstable when $\kz \rightarrow 0$. This can be seen from Eq.~(\ref{eq:kc3}) which in small $\kz$ limit gives
\bq
\left(\frac{k_c\,v_A}{\Omega}\right)^2 \sim \frac{1}{\Bz^2}\,.
\eq    
Therefore, $\kz\,k_c \rightarrow 0$ when $\kz \rightarrow 0$. Since fluctuations of arbitrary small wavelength can easily fit within the disc scale height (for right range of plasma beta), upper layer of the disc belonging to regions II and III in Hall—-ambipolar diffusion plane [Fig.~(\ref{fig:abr})] will be susceptible to MRI and DMRI in the presence of favourable field topology, i.e. when $\B\cdot\Omega < 0$ and $g > 0$. 

\subsection{MRI operates much closer to the midplane of PPDs when field is non--vertical.}
In the middle and midplane region of PPDs where Hall-Ohm diffusion is the dominant diffusion \citep{W07, B11}, the critical wavenumber is given by Eq.~(\ref{eq:kc1}). From this expression, we infer that in purely Hall dominated discs, waves propagating along the magnetic field, ($\kz = 1$) have the shortest wavelength. Indeed, setting $\eta_O =0$ in Eq.~(\ref{eq:kc1}), we see that $\kz\,k_c \rightarrow 0$ when $\kz \rightarrow 0$. Therefore, the shortest wavelength that fits within scale height is those that are propagating vertically. 

In the absence of Ohm diffusion, for $\kz = 1$, the critical wavenumber Eq.~(\ref{eq:kc1}) has following dependence on Hall diffusivity when field is almost vertical ($\Bz \rightarrow 1$)
\bq
\left(\frac{k_c\,v_A}{\Omega}\right)^2 \approx \frac{3}{2}\,\frac{1}{\left( \hetaH + 1/2 \right)}\,.
\label{eq:kcx}
\eq    

We have shown in section 3.1 that in Hall—-Ohm diffusion dominated discs general field topology with oblique wavevector can be mapped to pure vertical field and pure vertical wavenumber. What is the role of field topology and obliqueness of wavevector in such a disc? In order to answer this question we note from Eq.~(\ref{eq:kc1}) that when disc is threaded predominantly by azimuthal field i.e. $\Bz \rightarrow 0$ critical wavenumber can become arbitrarily large since $k_c^2 \propto 1 / \Bz$. However, no matter how small Ohm diffusion is, it can not be ignored when $\mu^2\,\eta_H^2$ term in the denominator of Eq.~(\ref{eq:kc1}) becomes comparable to $\eta_O^2$, i.e. $\Bz \approx \eta_O / \eta_H \sim 1 / \beta_e$. Thus for weak vertical field, critical wavelength becomes
\bq 
\left(\frac{k_c\,v_A}{\Omega}\right)^2 \approx \frac{3}{2}\,\frac{\Bz\,\hetaH}{\hetaO^2 + \mu^2\,\hetaH^2} \approx \frac{3}{4}\,\frac{1}{\hetaO}\,.
\label{eq:kcx1}
\eq
How small this vertical field can become? To answer this, we recall that the stability condition in a Hall—-Ohm dominated disc, Eq.~(\ref{eq:CH1}), for a dustless disc can be written as 
\bq
B_z > - 5.2 \,10^{-3} X_{e-8}^{-1}\,\Omega_{1\,\mbox{yr}}\,\,\mbox{G}\,.
\label{eq:CHC}
\eq
where $X_{e-8} = X_{e} / 10^{-8}$ and $ \Omega_{1\,\mbox{yr}} = \Omega / \ 1\,\mbox{yr}$.  Thus field can not become smaller than $\mbox{m}\,\mbox{G}$ or else Hall will cease to affect MRI. Clearly, coupling of the field to the disc matter sets the limit on the smallest vertical field. If the field is smaller than Eq.~({\ref{eq:CHC}), the field is not capable of exciting MRI in the disc. 

It is clear from Eqs.~(\ref{eq:kcx})-—(\ref{eq:kcx1}) that in Hall--Ohm diffusion dominated disc, when vertical field is weak, critical wavenumber scales with Ohm diffusion as $k_c^2 \propto 1 / \eta_O$. Recall that in a purely vertical field threaded disc $k_c^2 \propto 1 / \eta_O^2$ \citep{WS11}. This implies that MRI can operate for much larger values of Ohmic resistivity than is possible in a purely vertical field threaded disc. Therefore, we may infer that general field topology permits MRI to operate much closer to the midplane than is otherwise possible in discs threaded purely by vertical magnetic field.

Although our investigation is limited to the linear analysis, it is important to connect it to nonlinear investigation of the non—-ideal regimes in PPDs. The result of numerical simulation of PPDs in Hall—-Ohm regime \citep{SS02}  suggests that the quenching of turbulence continues to be set by Ohm diffusion even in the presence of Hall effect. However, as has been shown by WS11, \cite{SS02} assume that Ohm diffusion dominates Hall in their simulation. Therefore, it is not surprising that the results of \cite{SS02} shows that MRI turbulence is quenched by Ohmic diffusion. The effect of Hall diffusion on the saturation of the instability is yet to be investigated.  

\subsection{Purely vertical field and transverse fluctuation is fundamental to MRI in Hall—-Ohm dominated disc.}
The general field topology and obliqueness of wavevector merely rescales the growth rate and wavevector window. However, due to such scaling, fluctuations with smaller wavenumber becomes susceptible to MRI in azimuthal field dominated discs.  Although, discs can become MRI susceptible to broader fluctuation spectrum for increasingly weaker vertical field, no matter how small the vertical field is, its presence is necessary to excite the MRI. Discs with purely toroidal field are magnetorotationally stable to axisymmetric fluctuations. 

\subsection{PPDs are also susceptible to purely diffusive instability.}
PPDs are not only prone to MRI or, DMRI but can also be subject to shear driven diffusive instability. This is due to the fact that the magnetic field drift in a weakly ionised matter can occur not only due to advective fluid motion but can also be caused by the plasma—-neutral collision. As a result, in the high frequency limit, when field drift is solely due to the plasma-—neutral collision, waves in a differentially rotating disc can become unstable due to slippage of the field against the sea of neutrals–- the diffusion instability (DI). On the other hand, in the low frequency limit, when fluid advection is completely offset by the field diffusion, disc is prone to diffusive MRI (DMRI). The maximum growth rate of both DI and DMRI are similar. Since DI is high frequency counterpart of DMRI it implies that much before the onset of DMRI, the DI will destabilise the disc without significant transport of angular momentum. Since only weak turbulence is excited in excessively diffusive discs [Fig.~13, \cite{BS11}], DI may not impact the angular momentum transport due to DMRI.        

\section{Summary}

The MRI in a partially ionised diffusive disc has been analysed in the present work. Following is the brief summary of the work. 

1. The general stability criterion of a magnetised, differentially rotating diffusive disc [with a rotation profile $\Omega \propto r^{-q}\,,$ $0 < q < 2\,$] can be stated purely in terms of magnetic diffusivities as
\bq
 g\,\eta_A + s\,\eta_H > - \frac{2\,\mu^2\,v_A^2 / \Omega}{2 - q}\,.
\label{eq:spwd}
\eq
This criterion suggests that subthermal diffusive discs are always unstable to almost radial fluctuations.  

2. The MRI can operate much closer to the midplane in discs threaded by non--vertical magnetic field.  

3. Whereas discs which are Hall--Ohm stable  may or may not be ambipolar unstable, Hall--Ohm unstable discs  in the presence of non--vertical field and oblique wavevector can always become ambipolar unstable. 

4. Favourable field topology and oblique wavevector not only make disc magnetorotationally unstable in ambipolar regime, but simultaneously enlarges the domain over which Hall diffusion can assist MRI. 

5. The vertical magnetic field and transverse axisymmetric fluctuations (vertical wavevector) plays fundamental role in destabilising Hall--Ohm diffusion dominated disc. Furthermore, MRI grows at maximum rates in vertical field threaded discs. The non--vertical field and oblique wavevector can be mapped to a purely vertical field and vertical wavevector. 

6. The maximum MRI growth rate of the ambipolar diffusion assisted MRI is two fifth of ideal MHD case ($3/4\,\Omega$) for long wavelength fluctuations. Short wavelength fluctuations are likely to be damped by the ambipolar diffusion.      

7. Excessively diffusive discs can become unstable solely due to interplay between rotational shear of the disc and field diffusion. The ensuing diffusive instability grows at the same maximum rate as MRI. 

\section*{Acknowledgments}
{The financial support of Australian Research Council through grant DP0881066 is gratefully acknowledged.}

\label{lastpage}
\end{document}